\pdfoutput=1
\documentclass[prd,aps,twocolumn,preprintnumbers,nofootinbib,superscriptaddress,10pt]{revtex4-2}
\usepackage{graphicx}

\usepackage{silence}

\usepackage{amsmath,amsthm,amssymb,multirow}
\usepackage[normalem]{ulem}
\usepackage{color}
\usepackage{hyperref}
\usepackage{mathrsfs}
\graphicspath{{./Figures/}}

\begin{document}

\def\lsim{\mathrel{\rlap{\lower4pt\hbox{\hskip1pt$\sim$}}
  \raise1pt\hbox{$<$}}}
\def\gsim{\mathrel{\rlap{\lower4pt\hbox{\hskip1pt$\sim$}}
  \raise1pt\hbox{$>$}}}
\newcommand{\vev}[1]{ \left\langle {#1} \right\rangle }
\newcommand{\bra}[1]{ \langle {#1} | }
\newcommand{\ket}[1]{ | {#1} \rangle }
\newcommand{\ev}{ {\rm eV} }
\newcommand{\kev}{{\rm keV}}
\newcommand{\Mev}{{\rm MeV}}
\newcommand{\Gev}{{\rm GeV}}
\newcommand{\Tev}{{\rm TeV}}
\newcommand{\mw}{$M_{W}$}
\newcommand{\Ft}{F_{T}}
\newcommand{\Zparity}{\mathbb{Z}_2}
\newcommand{\BLambda}{\boldsymbol{\lambda}}
\newcommand{\met}{\;\not\!\!\!{E}_T}
\newcommand{\beq}{\begin{equation}}
\newcommand{\eeq}{\end{equation}}
\newcommand{\bea}{\begin{eqnarray}}
\newcommand{\eea}{\end{eqnarray}}
\newcommand{\nn}{\nonumber\\}
\newcommand{\gev}{{\mathrm GeV}}
\newcommand{\hc}{\mathrm{h.c.}}
\newcommand{\eps}{\epsilon}
\newcommand{\bwt}{\begin{widetext}}
\newcommand{\ewt}{\end{widetext}}
\newcommand{\draftnote}[1]{{\bf\color{blue} #1}}
\newcommand{\vecsigma}{ \mbox{\boldmath $\sigma$}}

\newcommand{\Mpl}{M_{\rm Pl}}
\newcommand{\TRH}{T_{\rm RH}}
\newcommand{\TCMB}{T_{\rm CMB}}
\newcommand{\xiend}{\xi_{\rm end}}
\newcommand{\Hend}{H_{\rm end}}
\newcommand{\HRH}{H_{\rm RH}}
\newcommand{\aRH}{a_{\rm RH}}
\newcommand{\aend}{a_{\rm end}}
\newcommand{\amax}{a_{\rm max}}
\newcommand{\abar}{a_{\rm bar}}
\newcommand{\kend}{k_{\rm end}}
\newcommand{\kmax}{k_{\rm max}}
\newcommand{\xend}{x_{\rm end}}
\newcommand{\kast}{k_\ast}
\newcommand{\rhoEend}{\rho^{\rm{end}}_E}
\newcommand{\rhoERH}{\rho^{\rm{RH}}_E}
\newcommand{\rhoIend}{\rho^{\rm{end}}_I}
\newcommand{\sigend}{\overline{\sigma}_{\rm{end}}}
\newcommand{\sigRH}{\overline{\sigma}_{\rm{RH}}}
\newcommand{\rhochiend}{\rho^{\rm{end}}_\chi}
\newcommand{\mbchi}{\overline{m}_\chi}
\newcommand{\sigb}{\overline{\sigma}}

\newcommand{\cO}{{\cal O}}
\newcommand{\cL}{{\cal L}}
\newcommand{\cM}{{\cal M}}

\newcommand{\be}{\begin{equation}}
\newcommand{\ee}{\end{equation}}

\newcommand{\fref}[1]{Fig.~\ref{fig:#1}} 
\newcommand{\eref}[1]{Eq.~\eqref{eq:#1}} 
\newcommand{\aref}[1]{Appendix~\ref{app:#1}}
\newcommand{\sref}[1]{Section~\ref{sec:#1}}
\newcommand{\tref}[1]{Table~\ref{tab:#1}}

\newcommand{\LU}[1]{\textcolor{blue}{[LU: #1]}}
\newcommand{\RVM}[1]{\textcolor{blue}{\bf{#1}}}

\title{\Large{{\bf Reheating effects on inflationary dark photon dark matter}}}

\author{\bf {Mar Bastero-Gil}}
\email{mbg@ugr.es}
\affiliation{\normalsize\it
Departamento de F\'{i}sica Te\'{o}rica y del Cosmos and CAFPE,~Universidad de Granada, Campus de Fuentenueva, E-18071 Granada, Spain}

\author{\bf{Paulo B. Ferraz}}
\email{paulo.ferraz@pi.infn.it}
\affiliation{\normalsize\it INFN, Sezione di Pisa, Largo B. Pontecorvo 3, 56127 Pisa, Italy}
\affiliation{\normalsize\it
Departamento de F\'{i}sica Te\'{o}rica y del Cosmos and CAFPE,~Universidad de Granada, Campus de Fuentenueva, E-18071 Granada, Spain}
\affiliation{\normalsize\it
Universidade de Coimbra, Faculdade de Ciências e Tecnologia da Universidade
de Coimbra and CFisUC, Rua Larga, 3004-516 Coimbra, Portugal}

\author{\bf{Pedro Garc\'{i}a Osorio}}
\email{pgosorio@correo.ugr.es}
\affiliation{\normalsize\it
Departamento de F\'{i}sica Te\'{o}rica y del Cosmos and CAFPE,~Universidad de Granada, Campus de Fuentenueva, E-18071 Granada, Spain}

\author{\bf{Lorenzo Ubaldi}}
\email{lorenzo.ubaldi@ijs.si}
\affiliation{\normalsize\it
Jo\v{z}ef Stefan Institute, Jamova 39, 1000 Ljubljana, Slovenia}
\affiliation{\normalsize\it
Institute for Fundamental Physics of the Universe (IFPU), \\ Via Beirut 2, 34014 Trieste, Italy}

\author{\bf{Roberto Vega-Morales}}
\email{rvegamorales@ugr.es}
\affiliation{\normalsize\it
Departamento de F\'{i}sica Te\'{o}rica y del Cosmos and CAFPE,~Universidad de Granada, Campus de Fuentenueva, E-18071 Granada, Spain}

\begin{abstract}
In this work we study the effects of the reheating phase on the inflationary dark photon dark matter production mechanism through an axial coupling of a dark $U(1)$ gauge field to the inflaton. The oscillations of the inflaton around the minimum of its potential induce parametric resonances that lead to additional production of certain dark photon modes, in some cases resulting in a larger dark matter abundance at the end of reheating than at the end of inflation. We examine the full dynamics during oscillations in the homogeneous backreaction picture employing the gradient expansion formalism. The final dark photon dark matter relic abundance, and consequently its expected mass range, depends not only on the duration of reheating, but also on the type of inflaton potential. We find the dark photons can be as light as $10^{-5}\,\text{eV}$ and as massive as $ 10^{7}\,\text{GeV}$ with a peaked power spectrum at small scales exhibiting oscillatory features which imprint the effects of reheating and the inflaton potential. 

\end{abstract}

\maketitle

\section{Introduction} \label{sec:intro} 

The nature and production mechanism of dark matter is still unknown and, as the `WIMP' paradigm has come into increasing tension over the years~\footnote{Despite the general lack of experimental evidence for WIMP dark matter during the last decade(s), the recent LZ Dark Matter Experiment \cite{LZ:2026axp} has presented results that are compatible with a WIMP-like dark matter candidate, but are not necessarily incompatible with non-thermal DM~\cite{Nagata:2026pbj}.}, candidates for non-thermal dark matter have gained renewed interest~\cite{Bertone:2018xtm}. One class of non-thermal dark matter which has received considerable attention in the recent past is inflationary-produced vector dark matter, for which a number of production mechanisms \cite{Nelson:2011sf, Arias:2012az, Graham:2015rva, Bastero-Gil:2018uel, Ema:2019yrd, Nakayama:2020rka, Nakai:2020cfw, Ahmed:2020fhc, Kolb:2020fwh, Salehian:2020asa, Firouzjahi:2020whk, Lysenko:2025sdo,Dror:2018pdh, Cline:2024wja,Cline:2026eps, Arza:2026rsl} have been constructed\footnote{A number of interesting mechanisms for producing vector dark matter after inflation have also been constructed \cite{Agrawal:2018vin,Co:2018lka}.}. 

One such mechanism arises from the Chern-Simons coupling between an axion-like inflaton and a dark $U(1)$ gauge field causing the gauge field to be produced exponentially \cite{Anber:2006xt,Anber:2009ua} through a process known as a tachyonic instability. When the dark gauge field obtains a mass, the resulting massive dark vector\footnote{We use `dark photon' and `dark vector' interchangeably throughout to refer to a neutral spin-one massive vector boson associated with a broken dark $U(1)_D$ gauge symmetry.} can serve as dark matter and reproduces the relic abundance in our universe over a very large range of masses \cite{Bastero-Gil:2018uel,Bastero-Gil:2021wsf}. 

These dark photons would already be present early during the slow-roll phase of inflation. However, since the dark photon energy density depends exponentially on the inflaton velocity, the most important contribution to dark matter production takes place at the end of inflation, just as the universe transitions from its inflationary state to a radiation dominated phase. Although it is possible to treat this reheating transition as instantaneous and assume the universe becomes radiation dominated the moment inflation ends (implying production of dark photons is fully halted at that point), a more realistic treatment must include the production during the reheating phase, in which the inflaton oscillates about its minimum\footnote{Lattice simulations in the context of magnetogenesis \cite{Cuissa:2018oiw,Figueroa:2023oxc,Figueroa:2024rkr,Iarygina:2025ncl}
have studied this regime in some range of the parameter
space.}. During this oscillatory phase, production of the dark photon is dominated by parametric resonance. The interplay between this additional production and redshifting during oscillations determines whether there is an increase or decrease in the dark photon energy density with respect to the end of inflation.

In order to obtain a more complete description of the possible dark matter parameter space, in this work we study the behavior of the dark photon production during reheating following the end of inflation, both in the case of a quick, instantaneous-like reheating and a slow, more inefficient reheating including an oscillatory phase for the inflaton. To this end, we employ numerical simulations in order to properly capture the evolution of the energy densities of the different components of the universe, in this case the inflaton, the dark photons and the visible (Standard Model) radiation that gives rise to the radiation-dominated era, all the way from the slow-roll phase to the end of reheating.

Due to the scale-dependent nature of this production mechanism, it is not trivial to find an accurate description of the process in terms of energy densities, as the resulting dark photon energy density (and its associated "dark" electromagnetic fields) will be largely non-homogeneous. As a partial solution to this problem we use the {\em homogeneous backreaction} description, which takes the spatial dependence of the dark photon distribution into account, but assumes the strength of the backreaction effects on the inflaton is small enough that the latter can still be assumed to remain homogeneous throughout all its evolution. Within this description, we use two methods to describe the dark photons' evolution. The first is a mode-by-mode integration of the Fourier modes, using integrals over momenta to obtain the total energy density and the electromagnetic fields. The second is the gradient expansion formalism, developed in \cite{Sobol:2019xls,Sobol:2021rlt}, which encodes the dark photons' spatial information in the form of an infinite tower of functions of the dark electromagnetic fields.

Using this formalism, we model the reheating phase with a constant decay rate of the inflaton into visible-sector degrees of freedom.
For a large decay rate the reheating happens very quickly and we find results qualitatively consistent
with those in a previous work~\cite{Bastero-Gil:2021wsf}, which assumed instantaneous reheating.
For a smaller decay rate the inflaton goes through oscillations, and we find that the results strongly depend on the type of inflaton potential, specifically on whether it behaves like a quadratic ($\phi^2$) or a quartic ($\phi^4$) monomial when expanded around the minimum. We also find a strong dependence on the reheating efficiency, as a more prolonged reheating phase can lead the dark photon energy density to increase (due to continued production) or decrease (due to redshift) by several orders of magnitude with respect to its value at the end of inflation. The shape of the resulting dark matter energy density spectrum also depends on reheating efficiency and the inflaton potential, showing distinct features due to parametric resonance.

This work is organized as follows: First, we review the details of the dark photon production model, showing the two methods we have used to describe their evolution. Next we describe how the inflaton's energy density is transferred to the radiation degrees of freedom during the reheating phase. We also discuss the behavior of the inflaton potential during the oscillatory phase including parametric resonance. We then present some numerical results for the evolution of the energy densities and the dark matter spectra, for different cases of the inflaton potential and the reheating efficiency. Afterwards, we study the cosmic evolution of the dark photons until present day and compare with the observed dark matter relic abundance in order to obtain the final parameter space for this model. Finally, we present our conclusions and prospects for future work.

\section{Production during inflation}\label{sec:darkphotonproduction}

Following \cite{Bastero-Gil:2018uel,Bastero-Gil:2021wsf}, we consider a dark matter model consisting of a massive vector boson associated to a broken abelian gauge symmetry $U(1)_D$. The dark photon is coupled to the inflaton field, $\phi$, via a Chern-Simons coupling.
This is typical of an axion field, but as we are 
interested in the dynamics of the 
production mechanism rather than in model building,
we need not necessarily assume an axion; 
any scalar with such a coupling will do.
In order to reheat the universe after the inflationary phase is over, the inflaton is coupled directly or indirectly to the Standard Model degrees of freedom. 

Our theory is described by the following action\footnote{Throughout this work we use natural units $c = \hbar = k_B = 1$ and consider a flat FLRW metric with the {\em mostly plus} sign convention, $ds^2 = -dt^2+a^2(t)\delta_{ij}dx^i dx^j$.},
\begin{eqnarray}
S&=-\int d^4 x \sqrt{-g}
\Big(\frac{1}{2}\partial_\mu\phi\partial^\mu\phi+V(\phi)+\frac{1}{4}F_{\mu\nu}F^{\mu\nu}+\nonumber\\
&+\frac{1}{2}m_A^2A_\mu A^\mu+\frac{\beta}{4m_P}\phi F_{\mu\nu}\tilde F^{\mu\nu}-\mathcal L_{\rm{SM}}\Big),
\end{eqnarray}
where $\phi$ is the inflaton field, $V(\phi)$ its potential, $\mathcal L_{SM}$ is the part of the Lagrangian containing the Standard Model (SM) degrees of freedom and their interactions with the inflaton, $A_\mu$ is the dark vector field, $F_{\mu\nu} = \partial_{[\mu}A_{\nu]}$ the gauge field strength, and $\tilde{F}^{\mu\nu}=\frac{1}{2}\varepsilon^{\mu\nu\rho\lambda}F_{\rho\lambda}$ with $\varepsilon^{\mu\nu\rho\lambda}$ the completely antisymmetric tensor.
We use the conventions: $\varepsilon^{0123} = +1 / \sqrt{-g}$, $F_{0i} = -a(t) E_i$, $F_{ij} = a^2(t) \varepsilon_{ijk} B_k$, with $E_i$ and $B_i$ the components of the electric and magnetic fields associated to the gauge field. 
The parameter $\beta$ is a constant parameterizing the strength of the Chern-Simons coupling, and $m_A$ is the dark photon mass. This mass may be of Higgs'ed or Stueckelberg-type, and it can be zero or non-zero during inflation. As long as it is smaller than the Hubble parameter during inflation, its effect is negligible in the production mechanism. Thus, we set $m_A \simeq 0$ during its dynamical evolution throughout inflation and reheating, implying the dark photons are relativistic at production. Only later in their  cosmic evolution as the dark photons redshift and their typical
momentum becomes comparable to $m_A$, does the mass become 
relevant. Since we assume there are no particles charged under the $U(1)_D$ symmetry\footnote{As shown in \cite{Bastero-Gil:2023htv,Bastero-Gil:2023mxm}, if there are dark charged particles present they can be produced via the Scwhinger effect and contribute to the dark matter relic abundance.}, the coupling  $\beta$ and the dark photon mass $m_A$ are the only two free parameters in the dark sector.

We consider a standard inflationary setting, consisting of a quasi-de Sitter phase where the inflaton slowly rolls down its potential $V(\phi)$, dominating the total energy density of the universe, followed by a reheating phase where the inflaton transfers its energy to the visible radiation degrees of freedom. Throughout its evolution the inflaton is governed by its equation of motion \cite{Bastero-Gil:2021wsf}:
\begin{equation}
    \ddot\phi + (3H+\Gamma_\phi)\dot \phi + \partial_\phi V(\phi)=- \frac{\beta}{4m_p} \langle F_{\mu\nu}\tilde{F}^{\mu\nu}\rangle\label{eq:inflaton}.
\end{equation}
Dots denote derivatives with respect to cosmic time, $H=\dot a/a$ is the Hubble scale and $\Gamma_\phi$ is the decay rate describing the energy transfer from the inflaton to the visible (SM) radiation degrees of freedom, which will be discussed in the following section. The brackets $\langle\cdot\rangle$ indicate the spatial average of the fields, which can be obtained by integrating over all Fourier modes. The Hubble scale can be calculated in terms of the total energy density of the universe using the second Friedman equation,
\begin{equation}
    H^2=\frac{1}{3m_P^2}\left(\frac{1}{2}\dot\phi^2+V(\phi)+\rho_R+\rho_A\right),
\end{equation}
where $m_P=(8\pi G)^{-1/2}\approx2.4\cdot10^{18}\,\text{GeV}$ is the reduced Planck mass, $\rho_R$ is the visible radiation energy density and $\rho_A$ the dark photon energy density. We consider inflation to end when the slow-roll parameter satisfies,
\begin{eqnarray}
    \eps_H\equiv-\frac{\dot H}{H^2}=1.
\end{eqnarray}

Expanding the dark gauge field in terms of its Fourier modes with momentum $\boldsymbol{k}$, the equations of motion for the transverse modes $\boldsymbol{A}_T = A_+\boldsymbol{\epsilon}_++A_-\boldsymbol{\epsilon}_-$, defined as $\boldsymbol{\epsilon}_\pm\cdot \boldsymbol{k}=0$, are, assuming $m_A\simeq 0$, given by \cite{Bastero-Gil:2021wsf,Anber:2009ua,Barnaby:2010vf},
\begin{equation}
\ddot A_\pm + H \dot A_\pm + \left( \frac{k^2}{a^2}\mp\frac{k}{a}\frac{\beta}{m_P}\dot\phi \right)A_\pm = 0\label{eq:TransvModes}.
\end{equation}
Writing \eref{TransvModes} in terms of conformal time $d\tau = dt/a$, 
\begin{equation}
\left[\frac{\partial^2}{\partial \tau^2}+k^2\mp2k\xi aH\right]A_\pm(\tau,k) = 0,
\end{equation}
where $\xi=\frac{\beta \dot\phi}{2m_PH}$ is known as the instability parameter. This can be seen as the equation of a harmonic oscillator with a time-dependent frequency $\omega_k$,
\begin{equation}
\left[\omega_k^2\right]_{\pm}=k^2\mp 2k \xi aH.\label{eq:frequency}
\end{equation}
Without loss of generality we consider $\dot\phi>0$ during inflation, implying that the positive polarization of the dark photon, $A_+$, undergoes a tachyonic instability when $\omega_k^2<0$ for modes that satisfy the condition,
\begin{equation}
k<k_h\equiv 2|\xi|aH,\label{eq:kh}
\end{equation}
where $k_h$ is the {\em horizon momentum} and represents the highest momentum that is amplified. Note that, as $k_h$ increases in time, more modes become tachyonically amplified, but it is not necessarily a monotonic function in time. If at some point it starts decreasing, the modes with $k>k_h$ that were already amplified begin to oscillate, though around an excited state rather than returning to the vacuum. Thus, in order to have a more accurate bookkeeping of the modes that were amplified, a better definition for $k_h$ is \cite{Sobol:2021rlt},
\begin{equation}
	k_h(t)=\max_{t'\leq t}{\left[2|\xi(t')|a(t')H(t')\right]}~.\label{eq:kh_2}
\end{equation}
With this definition, all modes that were amplified are taken into account at all times. Since $\omega^2_k$ is most negative at $k\sim |\xi|aH$ and typically $|\xi|\sim \mathcal O(1)$, it is expected that the power spectrum also peaks at scales around the comoving horizon. For modes outside the horizon, these add up coherently and have a very large occupation number due to the exponential enhancement of the tachyonic instability. Thus, these dark photons are well described by a \textit{dark} electromagnetic field. Furthermore, due to the amplification of a preferred helicity $A_+$, we are left with a helical dark electromagnetic field.

The total energy density of the dark photon field can be then calculated in terms of the spatial average of its associated electric and magnetic fields, $\boldsymbol{E}$ and $\boldsymbol{B}$, as $\rho_A =\left(\langle\boldsymbol{E}^2\rangle+\langle\boldsymbol{B}^2\rangle\right)/2$, which can be expressed in terms of the polarization modes as \cite{Sobol:2021rlt},
\begin{align}
    \left\langle \boldsymbol{E}^2\right\rangle&=\sum_{\lambda=\pm}\intop_0^{k_h}\frac{dk}{k} \frac{k^3}{2\pi^2 a^2}\left|\dot A_\lambda\right|^2,\label{eq:Esqrd}\\
    \left\langle \boldsymbol{B}^2\right\rangle&=\sum_{\lambda=\pm}\intop_0^{k_h}\frac{dk}{k} \frac{k^5}{2\pi^2 a^4}\left| A_\lambda\right|^2,\label{eq: Bsqrd}\\
    \left\langle \boldsymbol{E}\cdot \boldsymbol{B}\right\rangle&=-\sum_{\lambda=\pm}\intop_0^{k_h}\frac{dk}{k}\lambda \frac{k^4}{4\pi^2 a^3}\frac{d}{dt}\left| A_\lambda\right|^2,\label{eq:EdotB}
\end{align}
where we also included the object $\left\langle \boldsymbol{E}\cdot \boldsymbol{B}\right\rangle=-\langle F_{\mu\nu}\tilde{F}^{\mu\nu}\rangle/4 $ coming from the coupling term in \eqref{eq:inflaton}, and $k_h$ is the horizon momentum defined in Eq.~\eqref{eq:kh}.

In order to study the dark photon dark matter parameter space, it is necessary to obtain the time evolution of the quantities in \eqref{eq:Esqrd} and \eqref{eq: Bsqrd} and with them the total dark photon energy density up until the end of the reheating phase. A simple numerical approach, known as the {\em mode-by-mode} (MBM) solution, would be integrating equation \eqref{eq:TransvModes} in time for a sufficient amount of modes and then using these to obtain a numerical estimate of the integrals in \eqref{eq:Esqrd}, \eqref{eq: Bsqrd} and \eqref{eq:EdotB}. However, the presence of the term $\left\langle\boldsymbol{E}\cdot \boldsymbol{B}\right\rangle$ in the inflaton's equation of motion would make it necessary for the integral in \eqref{eq:EdotB} to be calculated at every step of the integration, resulting in extended computation times. Furthermore, as discussed before, the relevant modes are generally of the scale of the comoving horizon, $k\sim aH$, up to $\mathcal O(1)$ factors. This results in a shift in frequency of no less than 20 orders of magnitude from CMB scales to the end of inflation. This makes it even more computationally demanding to ensure the necessary range of modes is covered while keeping a reasonable resolution in momenta. 

An alternative method to calculate the evolution of the necessary electromagnetic fields is via a formalism known as the {\em gradient expansion}, discussed in detail in the following section. Although this formalism is significantly more efficient, it relies on approximations that require the horizon momentum $k_h$ to be growing in time and no longer hold after the end of inflation once the inflaton velocity reaches its global maximum and it begins oscillating with a decaying amplitude. However, from this point one can rely on the MBM solution, using the self-correction algorithm proposed in \cite{Sobol:2023gwk}, to obtain a full description of the system up until the end of reheating.

Both methods rely on the assumption that the inflaton behaves like a homogeneous classical field throughout inflation and reheating, as shown in Eq.~\eqref{eq:inflaton}, which requires backreaction effects to be small enough that they can be approximated by their spatial average. This is known as the {\em homogeneous backreaction} picture. In the regime of strong backreaction from the gauge fields the inflaton inhomogeneities must be accounted for, and a lattice treatment is necessary. Throughout this work we avoid such a regime by confining our study to cases in which the hierarchy $\rho_\phi \gg \rho_A$ is maintained at all times.

There is an additional constraint coming from parametric resonances induced by the $\phi F\tilde F$ coupling. Once the value of $\beta$ is large enough, parametric resonance induces not only additional production of dark photon modes during inflaton oscillations, but also non-zero momentum modes of the inflaton, potentially leading to the fragmentation of the inflaton field \cite{Adshead:2015pva}. Since this regime is outside the homogeneous backreaction picture, it is difficult to estimate without calculating it directly, in this case we can only rely on previous lattice simulations where it has been shown \cite {Figueroa:2024rkr,Cuissa:2018oiw} that these effects become relevant for couplings such that $\beta\gtrsim 12$ \footnote{These studies were carried out assuming an inflaton potential of the form $V(\phi)\propto\phi^2$, so this condition may not be exact for other types of potentials. However, we expect that for other potentials it should also be around $\beta\gtrsim\mathcal O(10)$.}.
For this reason, besides maintaining the hierarchy $\rho_\phi(t)\gg\rho_A(t)$ mentioned above, we will also limit our study to values of the dark photon coupling which satisfy $\beta< 12$.

\subsection{The Gradient Expansion formalism}\label{sec:GrExp}
As seen in the previous section, the scale-dependent mechanism of dark photon production in this model requires a description capable of capturing the necessary spatial information. A viable tool for this is the Gradient Expansion formalism \cite{Sobol:2019xls,Sobol:2020frh,Sobol:2021rlt}, which defines the following functions in terms of the electromagnetic fields:
\begin{align}
\mathcal E^{(n)}=\frac{1}{a^n}\left\langle \boldsymbol E \cdot (\nabla \times )^n \boldsymbol E \right \rangle,\\
\mathcal B^{(n)}=\frac{1}{a^n}\left\langle \boldsymbol B \cdot (\nabla \times )^n \boldsymbol B \right \rangle,\\
\mathcal G^{(n)}=-\frac{1}{a^n}\left\langle \boldsymbol E \cdot (\nabla \times )^n \boldsymbol B \right \rangle.
\end{align}
Here the brackets indicate the spatial average, so
these functions encode the information about the spatial dependence of the dark electromagnetic fields without being spatially dependent themselves. It is possible to write equations of motion for each of these Gradient Expansion Functions (GEFs)\footnote{Different formulations of the Gradient Expansion formalism can be found in the literature, often with different versions of the source terms in the equations of motion. In this work, we follow \cite{Sobol:2021rlt}, in the limit of zero Schwinger conductivity.}:
\begin{eqnarray}
    \dot{\mathcal{E}}^{(n)} +(n+4)H\mathcal{E}^{(n)}-4H\xi\mathcal{G}^{(n)}+2\mathcal{G}^{(n+1)}&=&S_{\mathcal{E}},\hspace{2mm}\label{eq:GEEoM-E} \\
    \dot{\mathcal{B}}^{(n)} +(n+4)H\mathcal{B}^{(n)}-2\mathcal{G}^{(n+1)}&=&S_{\mathcal{B}},\hspace{2mm} \label{eq:GEEoM-B} \\
    \dot{\mathcal{G}}^{(n)} +(n+4)H\mathcal{G}^{(n)}-\mathcal{E}^{(n+1)}\nonumber\\
+\mathcal{B}^{(n+1)} -2H\xi\mathcal{B}^{(n)}&=& S_{\mathcal{G}},\hspace{2mm}\label{eq:GEEoM-G}
\end{eqnarray}
where the terms on the right-hand side are known as {\em source terms}. These account for the fact that $k_h$, which acts like a cutoff to the Fourier modes that contribute to the integrals in \eref{Esqrd}, \eqref{eq: Bsqrd} and \eqref{eq:EdotB} (and their analogous for higher order GEFs), is increasing in time,
\begin{align}
    S_\mathcal{E}&=\frac{d\ln k_h}{dt} \frac{1}{4\pi^2} \left(\frac{k_h}{a}\right)^{n+4}\sum_{\lambda=\pm  1}\lambda^n E_\lambda (\xi),\\
    S_\mathcal{B}&=\frac{d\ln k_h}{dt} \frac{1}{4\pi^2} \left(\frac{k_h}{a}\right)^{n+4}\sum_{\lambda=\pm  1}\lambda^n B_\lambda (\xi),\\
    S_\mathcal{G}&=\frac{d\ln k_h}{dt} \frac{1}{4\pi^2} \left(\frac{k_h}{a}\right)^{n+4}\sum_{\lambda=\pm  1}\lambda^{n+1} G_\lambda (\xi).
\end{align}
The functions $E_\lambda$, $B_\lambda$ and $G_\lambda$ can be expressed in terms of Coulomb wave functions $F_l(n,\rho)$ and $G_l(n,\rho)$, and their derivatives $F_l'=\partial_\rho F_l$ and $G_l'=\partial_\rho G_l$:
\begin{align}
    E_\lambda(\xi)&=(G'_0)^2 +( F'_0)^2,\\
    B_\lambda(\xi)&=G_0^2+F_0^2,\\
    G_\lambda(\xi)&=-\left[G_0 G'_0+F_0 F'_0\right],
\end{align}
where $F_0\equiv F_0(\lambda \xi, 2|\xi|)$, and analogously for the rest \footnote{As Fortran lacks a native method for Coulomb wave functions, the results in this work have been obtained using the program COUL90 \cite{COUL90} which follows an algorithm described in \cite{CompAtomPhys}.}.

It is easy to notice in \eref{GEEoM-E}, (\ref{eq:GEEoM-B}) and (\ref{eq:GEEoM-G}) that the equations of motion of the GEFs of order $n$ only depend on the functions of the same order and the next. Therefore, by truncating the chain of equations at a certain $n=N_{\rm max}$ one obtains a set of $3 N_{\rm max}$ ODEs that can be integrated together with the inflaton's equation of motion, and in this way obtain the time evolution of the total electromagnetic energy density. Throughout this work we use the value $N_{\rm max}=90$, which we found to be large enough for results to be independent of $N_{\rm max}$ but small enough to avoid numerical issues when calculating the powers of $k_h/a$ in the source terms.

The main limitation of the Gradient Expansion comes from the condition used to cut off the expansion. When $N_{\rm max}$ is large, one may make the approximation \cite{Sobol:2021rlt},
\begin{equation}
\mathcal X^{(N_{\rm max}+1)}\approx\left(\frac{k_h}{a}\right)^2\mathcal X^{(N_{\rm max}-1)},\label{eq:cutoff}
\end{equation}
where $\mathcal X=\mathcal{E,B,G}$, and $k_h$ is defined in equation \eqref{eq:kh_2}. This approximation, however, only holds as long as $k_h$ is growing monotonously in time, that is, as long as its definition can be reduced to \eqref{eq:kh}. Since in that case $k_h(t)\sim a(t)\dot\phi(t)=d\phi/d\tau$, this is approximately equivalent to demanding that the inflaton is continuously accelerating. Therefore, approximation \eqref{eq:cutoff} can be used safely during the slow-roll phase and as the inflaton gains kinetic energy after the end of inflation. However, once the inflaton velocity begins to oscillate during reheating, this approximation no longer holds and the gradient expansion functions begin to gain an error that is independent of $N_{\rm max}$ and eventually propagates to the zero order functions \cite{Sobol:2021rlt}. At that point, using the gradient expansion is no longer viable, and in order to continue tracking the evolution of the electromagnetic fields one must switch over to the {\em mode-by-mode} solution, calculating the evolution of the dark vector polarization modes directly and integrating them using \eref{Esqrd}, \eqref{eq: Bsqrd} and \eqref{eq:EdotB} and using this information to periodically ''correct'' the gradient expansion functions, as proposed in the ''self-correction algorithm'' introduced in \cite{Sobol:2023gwk}.

This self-correction algorithm simply proposes that, once approximation \eqref{eq:cutoff} can no longer be used, one can check how much error the GEFs have gained with respect to the mode-by-mode solution:
\begin{equation}
\epsilon=\max_{X=\mathcal{E,B,G}}\left|\frac{X^{\rm GEF}-X^{\rm MBM}}{X^{\rm MBM}}\right|.
\end{equation}
Since only the zero-order GEFs are relevant for the dynamics, it is sufficient to evaluate the error for these at each step in time. Once the error surpasses a certain threshold, the mode-by-mode solution can be used to correct the gradient expansion by substituting the GEFs with their mode-by-mode values for all orders.

In the original version of this self-correction algorithm it was proposed to use a $1\%$ tolerance for the error and, when it is surpassed, correct the values of all GEFs for $n\geq 1$, in order to avoid small jumps of order $1\%$ in the values of the zero-order functions. However, in the present paper we have instead chosen a tolerance of $0.1\%$ for the error, and also chosen to substitute all functions including the $n=0$ ones, in order to reduce the deviation of the GEFs from the mode-by-mode solution as much as possible and obtain greater precision. The original algorithm also proposed comparing the values of the GEFs for two different values of $N_{\rm max}$ in order to check when the self-correction becomes necessary. For the sake of simplicity we begin correcting the GEFs with the mode-by-mode solution as soon as $k_h$ reaches its maximum value. 

As was already discussed earlier, one must take into account that, despite the fact that the gradient expansion is capable of accounting for the spatial dependence of the dark electromagnetic fields, it is still formulated in the homogeneous backreaction approximation and therefore assumes that the inflaton field's spatial dependence is negligible. Although together with the MBM solution it provides a complete description of the system within this picture (and it has been shown to correctly reproduce lattice results in the homogeneous backreaction regime \cite{Figueroa:2023oxc}), it stops being reliable as strong backreaction effects come into play.

\section{Production during reheating}\label{sec:reheating}

At the end of the inflationary period, the inflaton field starts the process of reheating the universe by transferring its energy to the visible radiation degrees of freedom, thus leading towards the radiation-dominated epoch that is the start of the thermal history of the universe.

In order to study the predictions of our model, it is first necessary to choose an inflaton potential $V(\phi)$ which largely determines the inflaton's dynamics throughout inflation and reheating. Although there exists a plethora of possible potentials, each with distinct behaviors at CMB scales and associated primordial spectrum predictions, the final parameter space of dark photon dark matter produced by this mechanism depends almost exclusively on the phenomena taking place at the end of inflation and throughout the oscillatory phase. As such, the results depend mainly on the leading-order term of the potential when it is expanded around its minimum at $\phi=0$,
\begin{equation}
    V(\phi) \approx \frac{\lambda_\phi}{p} m_P^4\left(\frac{\phi}{m_P}\right)^p,\label{eq: potential_mono}
\end{equation}
with $p=2$ for a quadratic potential and $p=4$ for a quartic potential. Many typical models such as chaotic inflation and axion inflation \cite{Freese:1990rb}, and Starobinsky inflation \cite{Starobinsky:1980te, Mijic:1986iv, Maeda:1987xf} take the approximation of Eq.~\eqref{eq: potential_mono} during the oscillatory phase \footnote{
Potentials which have an abrupt transition shortly before or after the end of inflation, such as in $\alpha$-attractor models, can't be approximated in this way and lead to distinct predictions than for the potentials considered here \cite{Gustavo}.}, with distinct values of $\lambda_\phi$ \footnote{Because we assume \eqref{eq: potential_mono} to be the first term in an expansion, we do not use $\lambda_\phi$ to fit the height of the potential to the scalar amplitude of the primordial power spectrum, $P_{\mathcal R}$.} and $n$ being associated to the different inflationary models.

The main focus of this work is the dynamics of the dark sector, which never thermalizes with the visible sector. In this spirit, to be as agnostic as possible on the details of reheating, we simply model the interaction between the inflaton and the visible sector with a generic constant perturbative decay rate $\Gamma_\phi$ \footnote{An exclusively perturbative description of this energy transfer implies the assumption that the underlying particle model is in the narrow resonance regime. Otherwise, non-perturbative effects would become highly relevant and require a complete description of the inflaton's perturbations, once again only realizable in the lattice \cite{Barman:2025lvk}.}, such that the energy density balance equations for the different components of the universe are,
\begin{align}
    &\dot\rho_{\phi} + 3H \dot{\phi}^2  = -\Gamma_\phi \dot\phi^2+\frac{\beta\dot\phi}{m_P}\left\langle \boldsymbol E \cdot \boldsymbol B\right\rangle\label{eq:inflatondensity},\\
    &\dot\rho_R+4H\rho_R=\Gamma_\phi \dot\phi^2\label{eq:radiation},\\
    &\dot\rho_A+4H\rho_A=S_A-\frac{\beta \dot\phi}{m_P}\left\langle \boldsymbol E \cdot \boldsymbol B\right\rangle\label{eq:rhoA},
\end{align}
where $\rho_\phi,$ $\rho_R$ and $\rho_A$ are the inflaton, visible sector and dark photon energy densities, respectively. We can obtain \eref{rhoA} as the sum of the zero-order cases of equations (\ref{eq:GEEoM-E}) and (\ref{eq:GEEoM-B}), substituting $\rho_A=\frac{1}{2}(\mathcal E^{(0)}+\mathcal B^{(0)})$. The source term $S_A\equiv\frac{1}{2}(S_\mathcal E+S_\mathcal B)$ accounts for the increase in the total energy density due to the fact that $k_h$, the upper bound in integrals in \eqref{eq:Esqrd} - \eqref{eq:EdotB}, increases during inflation. With this set of equations we can describe the evolution of the system throughout reheating.

We take $\Gamma_\phi$ to be zero during inflation, and we turn it on only at the end of it, which in principle corresponds to the moment when $\epsilon_H=1$. However, large values of $\Gamma_\phi$ at $\epsilon_H=1$, when the inflaton's potential energy is still larger than its kinetic energy, would quickly deplete the latter and effectively push the system back into the inflationary phase. In order to avoid this, we take a pragmatic approach and let the inflaton roll down its potential until it reaches $\epsilon_H = 2$, which occurs when $2V=\dot{\phi}^2/2$. For a potential of the form \eqref{eq: potential_mono} this occurs before the inflaton reaches, for the first time, its minimum at $\phi=0$. This choice lets us treat both the large and small $\Gamma_\phi$ cases, and allows for meaningful comparisons to the case of instantaneous reheating (for large $\Gamma_\phi$) considered in previous studies~\cite{Bastero-Gil:2018uel,Bastero-Gil:2021wsf}. Reheating ends once we have $\rho_\phi=\rho_R$ and after that point, the inflaton's energy density quickly dissipates as the visible radiation comes to dominate the universe.

There is an additional term in equations \eqref{eq:inflatondensity} and \eqref{eq:rhoA} which we have omitted: the contribution of the perturbative decay $\phi\rightarrow AA$ whose rate is given by,
\begin{equation}
	\Gamma_A=\frac{\beta^2}{64\pi}\frac{m_\phi^3}{m_P^2},\label{eq:GammaA}
\end{equation}
where $m_\phi^2=\partial_\phi^2V$ is the inflaton mass \cite{Adshead:2015pva,ParticleDataGroup:2014cgo}\footnote{Unlike $\Gamma_\phi$, we do not consider $\Gamma_A$ to be absent during inflation and then ''turned on'' during reheating, since its behavior is in principle fully determined by the interaction term in the Lagrangian and expression \eqref{eq:GammaA}.}. Since the dark vector does not interact with itself, we can simply treat this contribution as a separate component,
\begin{eqnarray}
    \dot\rho_{A,{\rm pert}}+4H\rho_{A,{\rm pert}}=\Gamma_A\dot\phi^2, \label{eq:rhoa_TH}
\end{eqnarray}
and impose $\rho_{A,{\rm pert}} \ll \rho_A$ as a condition to ensure it is negligible. This in practice puts a lower bound on $\beta$ to ensure non-perturbative production dominates over perturbative decays. We also neglect perturbative processes such as $A\phi\rightarrow A\phi$ which are expected to have a subdominant effect compared to $\phi \to AA$.

In order to avoid backreaction effects on the inflaton from perturbative decays into dark photons, we also impose a large hierarchy $\Gamma_\phi\gg\Gamma_A$ so that this decay channel can be neglected in the inflaton's equation of motion. In a quartic potential, the inflaton oscillations around $\phi =0$ result in $m_\phi\sim0$, and ensure that $\Gamma_A \ll \Gamma_\phi$. In the quadratic case the inflaton has a constant mass $m_\phi\simeq \sqrt{3}H_{\rm end}$, with $H_{\rm end}$ the value of the Hubble scale at the end of inflation. This gives a lower bound on the ratio $\Gamma_\phi/H_{\rm end}$:
\begin{eqnarray}
    \frac{\Gamma_\phi}{H_{\rm end}}\gg\frac{\Gamma_A}{H_{\rm end}}\approx  6.5\cdot10^{-11}\left(\frac{\beta}{12}\right)^2
    \left(\frac{ H_{\rm end}}{ 10^{13}\,\text{GeV}}\right)^2.
\end{eqnarray}
which, for $\beta=12$ (the largest value allowed by our formalism) and $H_{\rm end}=4.7\cdot10^{13}\,\text{GeV}$ (the largest value allowed by CMB constraints) gives $\Gamma_\phi/H_{\rm end}\gg1.4\cdot10^{-9}$, which we will use as a sufficient constraint from now on. Furthermore, given that the end of reheating, $\rho_\phi=\rho_R$, takes place approximately when $\Gamma_\phi$ becomes comparable to the Hubble rate, the hierarchy $\Gamma_\phi\gg \Gamma_A$ also implies $\Gamma_A\lesssim H$ at all times, implicitly protecting the model from thermalization of the dark sector with the inflaton and thus with the visible sector until the end of reheating.

\subsection{Parametric resonance}

While the inflaton oscillates around the minimum of its potential, parametric resonance effects come into play. Before turning to the numerical results, in this section we provide analytic estimates under the simplifying assumptions of no-backreaction in the equation of motion for the inflaton and small decay rate $\Gamma_\phi \ll H$ which only serves to damp the amplitude of the oscillations slowly. 

\begin{center}
    \emph{Quartic Potential}
\end{center}

Starting with the quartic case ($p=4$), it is convenient to use conformal time and the rescaled field $\chi\equiv a\phi$, in terms of which the Hubble friction disappears,
\begin{equation}
\chi''-\frac{a''}{a}\,\chi+\lambda_\phi\,\chi^3=0 .
\end{equation}
During this stage the energy density of the Universe is dominated by the oscillating inflaton, which on average behaves like radiation. Hence $a\propto\tau$, and $a''/a$ vanishes up to small oscillating corrections. The expansion then drops out of the equation entirely,
\begin{equation}
\chi''+\lambda_\phi\,\chi^3=0 ,\label{eq:chi4}
\end{equation}
a reflection of the classical conformal invariance of $\lambda_\phi\phi^4$. \eref{chi4} describes an anharmonic oscillator with conserved energy $\tfrac12\chi'^2+\tfrac14\lambda_\phi\chi^4=\tfrac14\lambda_\phi\chi_0^4$, where $\chi_0$ is the (constant) amplitude of the oscillations. Its exact solution is given by~\cite{Greene:1997fu},
\begin{equation}
\chi(\tau)=\chi_0\,{\rm cn}\!\left(\sqrt{\lambda_\phi}\,\chi_0\,(\tau-\tau_0)\,\middle|\,\tfrac12\right),\label{eq:cnsol}
\end{equation}
where ${\rm cn}(z|m)$ is the Jacobi elliptic cosine\footnote{${\rm cn}(z|m)\equiv\cos\vartheta$, where the angle $\vartheta$ is defined by $z=\int_0^{\vartheta}d\theta\,(1-m\sin^2\theta)^{-1/2}$. For $m=0$ it reduces to $\cos z$. It satisfies ${\rm cn}''=-(1-2m)\,{\rm cn}-2m\,{\rm cn}^3$, so for $m=\tfrac12$ it solves ${\rm cn}''=-{\rm cn}^3$, and it is periodic with period $4K(m)$, where $K(m)=\int_0^{\pi/2}d\theta\,(1-m\sin^2\theta)^{-1/2}$ and $K(\tfrac12)\simeq1.854$.}. Its period fixes the angular frequency of the oscillations,
\begin{equation}
\Omega=\frac{\pi}{2K(\frac12)}\,\sqrt{\lambda_\phi}\,\chi_0\simeq0.847\,\sqrt{\lambda_\phi}\,\chi_0 ,\label{eq:Omega4}
\end{equation}
which is constant in conformal time. The physical amplitude, $\Phi=\chi_0/a$, decays as $1/a$. The decay rate makes $\chi_0$, and hence $\Omega$, decrease slowly, as $e^{-\Gamma_\phi(t-t_\Gamma)/3}$, where $t_\Gamma$ is the time at which $\Gamma_\phi$ is switched on; for $\Gamma_\phi=10^{-3}H_{\rm end}$ this amounts to about $10\%$ over the whole reheating phase. Solving the background numerically we find $\chi_0\simeq1.1\,a_{\rm end}|\phi_{\rm end}|\simeq2.6\,a_{\rm end}m_P$ and $\Omega\simeq1.15\,a_{\rm end}H_{\rm end}$. These values are independent of $\lambda_\phi$, which only sets $H_{\rm end}$, and of $\beta$, as long as backreaction is negligible.
Although Eq.~\eqref{eq:cnsol} is not a harmonic function, it is very close to one. Its Fourier series contains only odd harmonics, with rapidly decreasing coefficients so that the approximate description,
\begin{equation}
\chi(\tau)\simeq\bar\chi\cos(\Omega\tau),\qquad\bar\chi\equiv0.955\,\chi_0 ,\label{eq:cosapprox}
\end{equation}
deviates from the exact solution by less than $5\%$ of $\chi_0$. 

Inserting $\phi'\simeq\chi'/a$, with $\chi$ from Eq.~\eqref{eq:cosapprox} (the corrections are of relative order $aH/\Omega$ which is small after the first oscillations), into~\eref{TransvModes} written in conformal time, and defining $2z=\Omega\tau-\pi/2$, we obtain
\begin{equation}
\left[\frac{\partial^2}{\partial z^2}+\Delta_k\pm2q\cos(2z)\right]A_\pm=0,\label{eq:Mathieu4}
\end{equation}
with the definitions,
\begin{equation}
\Delta_k=\frac{4k^2}{\Omega^2},\qquad q=2\beta\,\frac{k}{\Omega}\,\frac{\bar\chi}{a\,m_P}\ ,\label{eq:Mathieu4par}
\end{equation}
which gives \eref{Mathieu4} in the form of the Mathieu equation~\cite{Kofman:1997yn}. The sign in front of $q$ only shifts $z$ by $\pi/2$ so both helicities share the same stability chart: unstable, exponentially growing solutions exist whenever $(\Delta_k,q)$ lies inside one of the resonance bands. Three features of Eqs.~\eqref{eq:Mathieu4} and~\eqref{eq:Mathieu4par} determine the dynamics.

\begin{enumerate}
    \item 
 The frequency of a mode turns imaginary during part of each oscillation whenever $\Delta_k<2q$, i.e.\ for $k< \beta\,\Omega\,\bar\chi/(a\,m_P)$. This is the broad resonance regime, $q\gtrsim1$, in which the modes are amplified in bursts at each half-oscillation of the inflaton, alternating between the two helicities. As $q$ decreases, each mode eventually leaves this regime, and only narrow resonance bands remain. Both regimes are contained in the linear mode equation and are therefore captured by our mode-by-mode integration; the Mathieu analysis serves as a guide to interpret the numerical results to be discussed below.

\item
Since $\Delta_k$ does not depend on time, each mode keeps a fixed $\Delta_k$ while $q$ decreases as $1/a$. In the narrow regime, $|q|\ll1$, the bands are centred at $\Delta_k=n^2$, i.e.\ at the fixed comoving momenta
\begin{equation}
k_n=\frac n2\,\Omega\simeq0.57\,n\,a_{\rm end}H_{\rm end},\qquad n\in\mathbb N ,\label{eq:res_phi4}
\end{equation}
and the modes sitting at the centre of a band are amplified for as long as the inflaton oscillates. At the centre of the $n$-th band, Eq.~\eqref{eq:Mathieu4par} gives $q_n=n\beta\bar\chi/(a\,m_P)\simeq2.5\,n\beta\,e^{-N}$. The first band therefore becomes narrow ($q_1<1$) after $N\simeq\ln(2.5\beta)$, i.e.\ $N\simeq2.3$ for $\beta=4$. 
For $\Gamma_\phi=10^{-3}H_{\rm end}$ reheating ends at $N\simeq3.5$, so for $\beta=4$ the last e-fold or so is in the narrow regime. 

\item
In the narrow regime the amplitude of the mode at the centre of the first band grows at the rate $\mu=|q_1|\Omega/4$ in conformal time~\cite{Kofman:1997yn}, i.e.\ $|A|\propto\exp(\int\mu\,d\tau)$. Since $dN=aH\,d\tau$, the number of e-folds of growth of the amplitude per e-fold is,
\begin{equation}
\frac{d\ln|A|}{dN}=\frac{\mu}{aH}=\frac{|q_1|}{4}\,\frac{\Omega}{aH} .
\end{equation}
At the centre of the first band, $q_1=\beta\bar\chi/(a\,m_P)=0.955\,\beta\,\Phi/m_P$, with $\Phi=\chi_0/a$ the physical amplitude. While the inflaton dominates the energy density, we have $H=\sqrt{\lambda_\phi}\,\Phi^2/(2\sqrt3\,m_P)$, and from Eq.~\eqref{eq:Omega4} the physical frequency is $\Omega/a=0.847\sqrt{\lambda_\phi}\,\Phi$. Hence $\Omega/(aH)\simeq2.93\,m_P/\Phi$ and,
\begin{equation}
\frac{\mu}{aH}\simeq0.7\,\beta .
\end{equation}
The growth per e-fold is constant because $q_1\propto1/a$ while $\Omega/(aH)\propto a$. Including the radiation, which increases $H$, the numerical background gives $\simeq0.65\,\beta$. The mode at the centre therefore grows as $|A|\propto a^{0.65\beta}$, and since the energy of a mode scales as $|A|^2/a^4$, the spectral energy density at the centre of the band, $d\rho_A/d\ln k|_{k_1}$, scales as $a^{1.3\beta-4}$. It grows in absolute terms for $\beta\gtrsim3$, and relative to the inflaton ($\rho_\phi\propto a^{-4}$) for any $\beta\gtrsim1$. This is the origin of the sustained production observed for the quartic potential in \fref{mechanisms} and \fref{rho_evolution}. Higher bands are weaker: the band at $n=2$ ($k=\Omega$) is of second order in $q$, while the band at $n=3$ ($k=3\Omega/2$) receives a first-order contribution from the third harmonic of the pump, with a growth rate about $13.5\%$ of that of the first band.

\end{enumerate}

Thus, for long enough reheating, we expect the dark photon spectrum to develop a peak at $k_1$, and weaker features at $k_2$ and $k_3$, on top of the spectrum generated at the end of inflation. This does not erase the information about the tachyonic production which took place during inflation: in the linear regime, each mode at the end of reheating equals the mode at the end of inflation multiplied by a transfer factor that depends only on the background. Since the tachyonically produced modes all emerge with the same phase, this factor acts multiplicatively on the spectrum, and it can be computed from Eq.~\eqref{eq:Mathieu4} once the background is known.

\begin{center}
    \emph{Quadratic Potential}
\end{center}

The same analysis can be carried out for the quadratic case ($p=2$), with $m_\phi^2=\lambda_\phi m_P^2$. Here it is convenient to work in cosmic time. During the oscillations the inflaton dominates the energy density and behaves on average like matter, $\dot H\simeq-\tfrac32H^2$. With the rescaled field $\chi\equiv a^{3/2}\phi$, again neglecting backreaction and the decay rate, the inflaton equation of motion becomes,
\begin{equation}
\ddot\chi+\Big(m_\phi^2-\tfrac94H^2-\tfrac32\dot H\Big)\chi=\ddot\chi+m_\phi^2\chi=0 .
\end{equation}
The solution is an exact harmonic oscillator with decaying amplitude,
\begin{equation}
\phi(t)=\Phi(t)\cos(m_\phi t),\qquad \Phi=\chi_0\,a^{-3/2} .\label{eq:phi2sol}
\end{equation}
The decay rate adds a slow damping, $\Phi\propto a^{-3/2}e^{-\Gamma_\phi(t-t_\Gamma)/2}$. The numerical background gives $\Phi\simeq m_P\,(a_{\rm end}/a)^{3/2}$ to within about $10\%$ after the first oscillation, i.e.\ $\chi_0\simeq|\phi_{\rm end}|\,a_{\rm end}^{3/2}$ with $|\phi_{\rm end}|\simeq m_P$.

The mode equation~\eqref{eq:TransvModes} contains a friction term $H\dot A_\pm$, which is removed by the rescaling $A_\pm=a^{-1/2}\mathcal A_\pm$\footnote{This rescaling has no physical consequence for the amplification. For a freely oscillating mode, the adiabatic amplitude of $\mathcal A$ grows as $(k/a)^{-1/2}\propto a^{1/2}$, which exactly compensates the factor $a^{-1/2}$, so that $A$ has a constant amplitude, as expected for a conformal field. Amplification is measured by the Floquet growth of $\mathcal A$, or equivalently by the occupation number of the mode, neither of which is affected.}. In a matter dominated universe this gives,
\begin{equation}
\ddot{\mathcal A}_\pm+\left(\frac{k^2}{a^2}+\frac{H^2}{2}\mp\frac{k}{a}\frac{\beta}{m_P}\dot\phi\right)\mathcal A_\pm=0 .
\end{equation}
Defining $2z=m_\phi t-\pi/2$ and using Eq.~\eqref{eq:phi2sol}, again leads to the Mathieu equation~\eqref{eq:Mathieu4},
\begin{equation}
\left[\frac{\partial^2}{\partial z^2}+\Delta_k\pm2q\cos(2z)\right]\mathcal A_\pm=0,
\end{equation}
now with the definitions,
\begin{equation}
\Delta_k=\frac{4k^2}{a^2m_\phi^2},\qquad q=2\beta\,\frac{k}{a\,m_\phi}\,\frac{\Phi}{m_P}\ \propto\ \frac{k}{a^{5/2}} .\label{eq:Mathieu2par}
\end{equation}
Here we dropped the term $2H^2/m_\phi^2$ in $\Delta_k$ as well as the term $-\tfrac32H\Phi\cos(m_\phi t)$ in $\dot\phi$ which comes from the time dependence of $\Phi$, as both are suppressed by $H/m_\phi$. More importantly, $\Delta_k\propto a^{-2}$ and $q\propto a^{-5/2}$ depend on time, and treating them as constant, as Floquet theory requires, assumes that they change little during one oscillation. Over one period, $2\pi/m_\phi$, the scale factor grows by $\delta\ln a\simeq2\pi H/m_\phi$, so $\Delta_k$ and $q$ change by the fractions $\simeq4\pi H/m_\phi$ and $\simeq5\pi H/m_\phi$, respectively. Since $H/m_\phi\simeq\tfrac12e^{-3N/2}$, these changes are of order one during the first e-fold after inflation (at $N\simeq1.2$, $H/m_\phi\simeq0.08$ and $\Delta_k$ still changes by about $100\%$ per period) and become small only for $N\gtrsim2$. During the first two e-folds the resonance is thus not a sequence of well-defined Floquet bands but rather a stochastic resonance~\cite{Kofman:1997yn}. The Mathieu description below should therefore be understood as a qualitative guide, which becomes quantitative only later, to better understand the numerical results for which we solve the full mode equation.

The bands are centered at a fixed \emph{physical} momentum,
\begin{equation}
\frac{k_n}{a}=\frac n2\,m_\phi ,\qquad n\in\mathbb N ,\label{eq:res_phi2}
\end{equation}
so in comoving momentum they move towards the ultraviolet, $k_n\propto a$. With $m_\phi\simeq2H_{\rm end}$, the first band sits at $k_1\simeq(a/a_{\rm end})\,a_{\rm end}H_{\rm end}$, and it sweeps through the spectrum produced during inflation. At the center of the $n$-th band Eq.~\eqref{eq:Mathieu2par} gives,
\begin{equation}
q_n=n\beta\,\frac{\Phi}{m_P}\simeq n\beta\,e^{-3N/2},
\end{equation}
which decreases much faster than in the quartic case. For a given mode $k$, three stages can be distinguished.

\emph{Broad resonance, $N\lesssim\tfrac23\ln\beta$.} For $q_1\gtrsim1$, i.e.\ during the first $\simeq1.2$ e-folds for $\beta=6$, a wide range of modes, including those at the peak of the spectrum, $k\sim a_{\rm end}H_{\rm end}$, have $\Delta_k<2q$. Their frequency turns imaginary during part of each oscillation, and they are amplified in bursts at each half-oscillation of the inflaton, alternating between the two helicities. This stage is responsible for most of the dark photon production after inflation in the quadratic case, i.e.\ for the first 1-2 e-folds in \fref{mechanisms}.

\emph{Crossing of the narrow band.} Once $q_1<1$, a mode is amplified only while the first band sweeps through it, which happens when $k_1(t)\simeq k$, i.e.\ at $a/a_{\rm end}\simeq k/(a_{\rm end}H_{\rm end})$. Since $\Delta_k\propto a^{-2}$, the mode spends only $\Delta N\simeq q$ e-folds inside the band, $1-q\lesssim\Delta_k\lesssim1+q$. During this time the mode completes $q\,m_\phi/(2\pi H)\simeq\beta/\pi$ oscillations. To estimate the amplification we can compute the growth exponent accumulated during the crossing. Integrating the Mathieu growth rate, $\mu=\tfrac{m_\phi}{4}\sqrt{q^2-(\Delta_k-1)^2}$, across the band gives
\begin{equation}
G=\int\mu\,dt=\frac{\pi\,q^2\,m_\phi}{16\,H}\bigg|_{\rm crossing}\simeq\frac{\pi}{8}\,\beta^2\left(\frac{k}{a_{\rm end}H_{\rm end}}\right)^{-3/2} ,\label{eq:Gk}
\end{equation}
where in the last step we used $q\simeq\beta e^{-3N/2}$ and $m_\phi/H\simeq2e^{3N/2}$. This expression holds for $q<1$, i.e.\ for $k/(a_{\rm end}H_{\rm end})\gtrsim\beta^{2/3}$. Right after the resonance becomes narrow, $G\simeq\pi\beta/8$, so the amplification is not negligible. However, $G$ decreases as $k^{-3/2}$, and it falls below unity for $k/(a_{\rm end}H_{\rm end})\gtrsim(\pi\beta^2/8)^{2/3}$. Narrow-band crossings therefore leave visible features only in a finite window,
\begin{equation}\label{eq:bandcrossing}
\beta^{2/3}\lesssim\frac{k}{a_{\rm end}H_{\rm end}}\lesssim\Big(\frac{\pi\beta^2}{8}\Big)^{2/3} ,
\end{equation}
i.e.\ $3\lesssim k/(a_{\rm end}H_{\rm end})\lesssim6$ for $\beta=6$, around and above the peak of the spectrum. The modes in this window cross the band during $\tfrac23\ln\beta\lesssim N\lesssim\tfrac23\ln(\pi\beta^2/8)$, i.e.\ $1.2\lesssim N\lesssim1.8$ for $\beta=6$. Since each crossing lasts only $\sim\beta/\pi$ oscillations, and the crossings happen when $H/m_\phi$ is not yet small, the passage is far from the adiabatic limit in which Eq.~\eqref{eq:Gk} is exact. $G$ should therefore be taken as an order-of-magnitude estimate.

\emph{The band leaves the populated range.} After inflation $k_h$, Eq.~\eqref{eq:kh_2}, remains close to its value at the end of inflation, $k_h\simeq\sqrt2\,\beta\,a_{\rm end}H_{\rm end}$. The first band reaches it after,
\begin{equation}
N_1\simeq\ln\left(\sqrt2\,\beta\right),\label{eq:Ne_res_exit}
\end{equation}
i.e.\ $N_1\simeq2.1$ for $\beta=6$; since the actual $k_h$ is slightly larger, these values are lower limits. From then on, the band only acts on modes that were never tachyonically amplified and are essentially in the vacuum. These modes are not included in the mode integrals, which extend up to $k_h$. For the values of $\beta$ we consider, $G$ has already dropped below unity by the time $k_1$ reaches $k_h$.

In summary, for both potentials the oscillating inflaton drives the gauge modes through the same Mathieu equation, first in the broad (tachyonic) regime and then in narrow resonance bands. For $\phi^2$ the bands sweep through the comoving spectrum. The modes near the peak are amplified in the broad regime during the first e-fold, the modes crossing the first band shortly afterwards receive a finite amplification, Eq.~\eqref{eq:Gk}, and once $k_1>k_h$ only vacuum modes are affected. We therefore expect features in a limited range of momenta around and above the peak of the spectrum, and none at larger $k$. For $\phi^4$ the bands are fixed in comoving momentum, and the modes at their centres are amplified at a constant rate per e-fold for as long as the inflaton oscillates. We therefore expect growing peaks at $k_n$, Eq.~\eqref{eq:res_phi4}, for long reheating.

\subsection{Numerical analysis}

Using numerical simulations, we study the behavior of the dark photons during the oscillatory phase until the end of reheating. We compare the results for $V\sim\phi^2$ and $V\sim\phi^4$, as well as different values of the ratio $\Gamma_\phi/H_{\rm end}$ which we take as a measure of reheating efficiency. 

An oscillator under a potential of the form $V(\phi)\propto \phi^{p}$ will have its average kinetic and potential energy follow the relation given by the virial theorem,
\begin{eqnarray}
    \frac{1}{2}\langle\dot\phi^2 \rangle=\frac{p}{2}\langle V\rangle=\frac{p}{2+p}\rho_\phi.\label{eq:virial}
\end{eqnarray}
That is, the inflaton's kinetic energy represents half of its energy density in the quadratic case ($p=2$), but two thirds of the total in the quartic case ($p=4$). Using \eqref{eq:virial} to estimate the instability parameter, which serves as a measure of dark photon production, we find,
\begin{eqnarray}
\langle|\xi|\rangle=\frac{\beta}{m_P}\frac{\langle|\dot\phi|\rangle}{2H}
\sim
\sqrt{\frac{3p}{2(2+p)}}\beta,
\end{eqnarray}
during the stage where the inflaton still dominates the total energy density, and therefore $H^2\approx\rho_\phi/3m_P^2$. This gives $\langle\xi\rangle\sim 0.866\,\beta$ in the quadratic case and $\langle\xi\rangle\sim\beta$ in the quartic. Thus, in order to obtain the same amount of production of the dark photon (which depends exponentially on $\xi$), throughout the examples in this section we use $\beta=6$ for the quadratic case and $\beta=4$ for the quartic case since this choice yields a similar magnitude for the energy density in both cases.

In \fref{mechanisms} we observe the behavior of the different terms in equation \eqref{eq:rhoA} as inflation ends and the inflaton starts to oscillate. We use values of the inflaton self-coupling $\lambda_\phi$ such that $H_{\rm end}\simeq10^{13}\,\text{GeV}$ in both potentials, and take the limit $\Gamma_\phi\rightarrow0$ in order to observe only the effects derived from the oscillations of the inflaton. The term $S_A$, the contribution induced by the fact that the cutoff momentum $k_h$ in the mode integrals increases with time, is the dominant source of dark photon production during inflation. It compensates the Hubble dilution term, but disappears as inflation ends and the horizon momentum $k_h$ reaches its maximum (constant) value. After this, it is the coupling term proportional to $\langle\boldsymbol{E}\cdot\boldsymbol{B}\rangle$ which takes over, competing with Hubble dilution and continuing dark photon production during oscillations. 
\begin{figure*}[t]
    \centering
\includegraphics{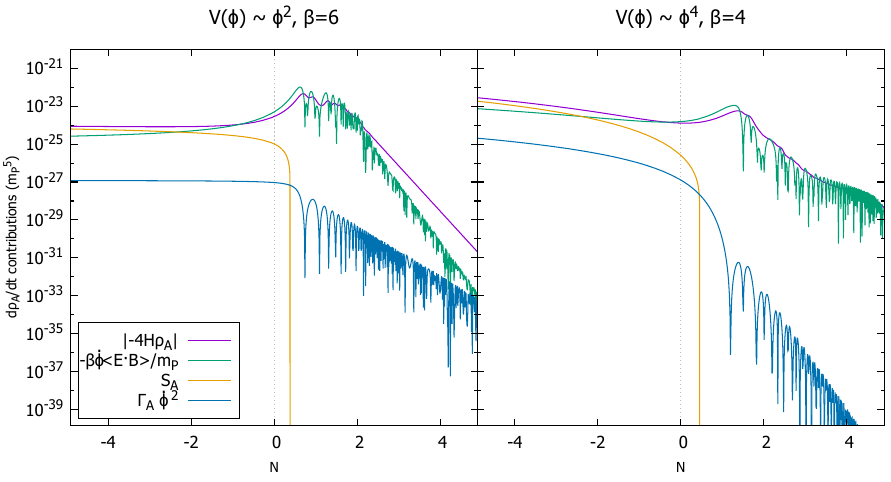}
    \caption{Comparison of the different contributions to the dark photon energy density's equation of motion,  \eqref{eq:rhoA}, along with the perturbative production term from \eqref{eq:rhoa_TH} with respect to the number of e-folds. These results correspond to $H_{\rm end} \simeq 10^{13}\,\text{GeV}$ and $\Gamma_\phi=0$. The dotted line at $N=0$ marks the end of inflation.}
    \label{fig:mechanisms}
\end{figure*}

The behavior of the coupling term presents the main difference between the two types of potential. In the quadratic case, dark photons continue to be produced during the first two e-folds after the end of inflation, but as gauge field production diminishes, the dilution term starts to dominate causing $\rho_A$ to redshift like radiation until the end of reheating. In the quartic case, however, although the coupling term is initially smaller, it stays comparable to the dilution term and dark photons continue to be produced as long as the inflaton oscillates. As can be seen in \fref{rho_evolution}, this leads to $\rho_A$ staying approximately constant as the other components of the universe redshift, eventually breaking the $\rho_\phi\gg\rho_A$ condition (for these parameters, this takes place at approximately $N\simeq5$). Thus, when using a quartic potential in this description, it is necessary to ensure we reach the end of reheating before this condition is broken. This behavior and difference in dark photon production among the two potentials can be understood in terms of the parametric resonances analysis presented in the previous subsection. In a $\phi^2$ potential with a coupling $\beta=6$, one can derive from \eqref{eq:Ne_res_exit} that the center of the first band $k_1$ becomes larger than $k_h$ precisely $N\simeq2$ e-folds after the end of inflation, causing resonant production to stop. In the quartic case, however, resonance bands are fixed at the same values of the comoving momentum $k$, and continue to contribute to non-vacuum modes as long as the inflaton oscillations continue.

The term representing the contribution of the perturbative decay rate $\Gamma_A$ defined in \eqref{eq:GammaA}, although shown to be negligible in general, also behaves differently for the two potentials due to the difference in the inflaton mass $m_\phi$. In the $\phi^4$ case, this mass oscillates around $m_\phi=0$, causing this term, which was already subdominant during slow-roll, to quickly decay after the oscillations start. In the $\phi^2$ case, however, the mass $m_\phi$ is constant, causing this term to decay at a slower rate and allowing it to eventually catch up with the other terms in the equation of motion. This shows that, in reality, both potentials have a limit to the duration of reheating in our formalism: In the $\phi^2$ case, we must ensure reheating ends before the perturbative contribution to the energy transfer becomes relevant in the equation of motion for $\rho_A$, while in the $\phi^4$ case the $\rho_\phi\gg\rho_A$ condition is broken unless reheating ends before the dark photons catch up with the inflaton's energy density.

Additionally, much like the gauge preheating effect induced by the $\phi F\tilde F$ coupling \cite{Adshead:2015pva}, the self-interaction of the inflaton introduced by the quartic potential may also cause non-linear parametric resonances among the inflaton's fluctuations themselves, potentially leading to the preheating of these fluctuations and an additional depletion of the inflaton's background energy. This effect cannot be calculated outside the lattice either. However, by looking at previous results \cite{Khlebnikov:1996mc} it is possible to extrapolate that these effects would take approximately 4 to 5 e-folds of oscillations to become strong enough to be relevant, giving yet another reason why our formalism cannot delay the end of reheating beyond 4 to 5 e-folds after the end of inflation.

\begin{figure*}[t]
    \centering
    \includegraphics[width=0.6\textwidth]{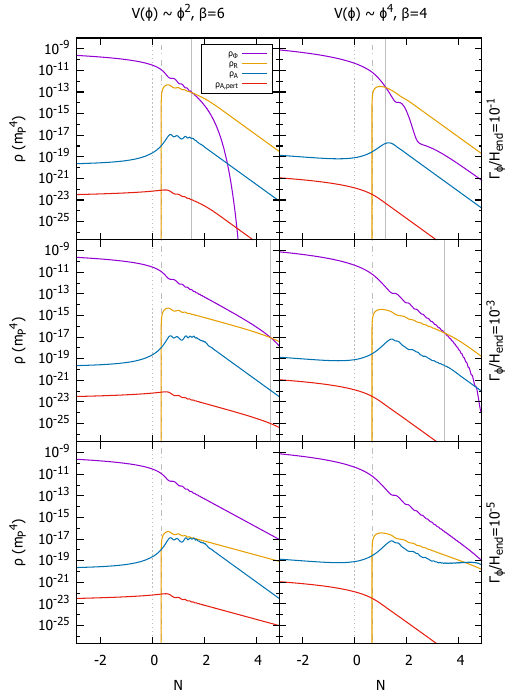}
    \caption{Evolution of the energy densities of the different components of the universe with respect to the number of e-folds. We consider a quadratic (first column) and quartic potential (second column) with $\beta=6$ and $\beta = 4$ respectively for distinct values of the ratio $\Gamma_\phi/H_{\rm end}$. These results correspond to $H_{\rm end} \simeq 10^{13}\,\text{GeV}$. The vertical dotted lines at $N=0$ mark the end of inflation ($\epsilon_H=1$), the dot-dashed lines mark the beginning of reheating (taken as $\epsilon_H=2$), and the solid lines mark the end of reheating defined as $\rho_\phi=\rho_R$. It is worth nothing that the plots in the bottom row break the assumptions that make our formalism reliable before the end of reheating is reached.}
    \label{fig:rho_evolution}
\end{figure*}

In Fig.~\ref{fig:rho_evolution} we show the evolution of the energy densities for the different components of the universe during the oscillatory phase using the same parameters as in \fref{mechanisms}, but now with three different values of the decay rate: $\Gamma_\phi  = 10^{-1}H_{\rm end}$ (top), $\Gamma_\phi=10^{-3}H_{\rm end}$ (center) and $\Gamma_\phi = 10^{-5}H_{\rm end}$ (bottom), with $H_{\rm end}\simeq 10^{13}\textrm{ GeV}$. As expected from the behavior observed in \fref{mechanisms}, looking at \fref{rho_evolution} we see that for the quadratic potential (first column), dark photons are continuously produced for approximately 1 to 2 e-folds past the end of inflation (that is, up until all resonance bands enter the horizon and resonant production ceases) causing their energy density to oscillate around a constant value. After this point, the dark photons stop being produced and simply begin to redshift like radiation ($\rho_A\sim a^{-4}$) and therefore faster than the rest of the components of the universe (for a quadratic potential $\rho_\phi\sim a^{-3}$ and $\rho_R\sim a^{-3/2}$ during the oscillatory phase). Due to this, the longer reheating is prolonged, the smaller the dark photon energy density will be in relation to the other components of the universe at the end of reheating.

For a quartic potential (second column), the behavior of the dark photon energy density changes considerably. Firstly, the inflaton starts behaving as a radiation fluid ($\rho_\phi\sim a^{-4}$) once its starts oscillating around the minimum of its potential, while the radiation redshifts as $\rho_R\sim a^{-2}$. This leads to $\rho_R/\rho_\phi\sim a^{2}$ which causes the reheating phase to be shorter in number of e-folds than in the quadratic case, where $\rho_R/\rho_\phi\sim a^{3/2}$. As for the dark photons, we observe that they are not produced as strongly during the first e-folds of reheating and then production decays after reaching a maximum value. However, if the decay rate is small and the oscillations continue, dark photon production will eventually pick up again, as seen in \fref{mechanisms}, and grow gradually stronger until the end of reheating. This leads to $\rho_A$ eventually becoming comparable to $\rho_\phi$ and $\rho_R$ if reheating takes long enough, breaking the $\rho_\phi\gg\rho_A$ assumption and transitioning into a strong backreaction regime, rendering our formalism unreliable from that point on.

\begin{figure*}[t]
\centering
\includegraphics[width = 0.7\textwidth]{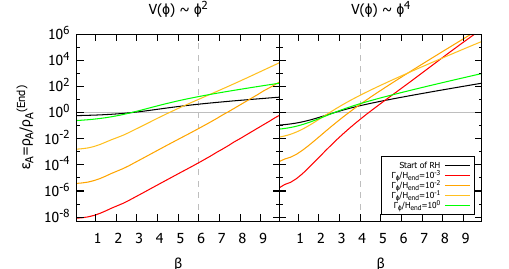}
    \caption{Plot of $\epsilon_A=\rho_A/\rho_A^{(\rm end)}$ with respect to the coupling $\beta$, evaluated at the start of reheating (black line) and the end of reheating for different values of the decay rate (colored lines). The horizontal line marks $\eps_A=1$ while the dashed vertical lines mark the values of $\beta$ used for each potential in Figs.~ \ref{fig:mechanisms} and \ref{fig:rho_evolution}. These curves were obtained using $H_{\rm end}\simeq10^9\,\text{GeV}$. Although the shape of the curves does not depend on the value of $H_{\rm end}$ as long as the strong backreaction regime is avoided, there may be values of $H_{\rm end}$ for which parts of these curves fall in the strong backreaction regime and therefore not match the results on this plot.}
    \label{fig:epsa}
\end{figure*}
The perturbatively produced contribution to the dark photon energy density, $\rho_{A,{\rm pert}}$, is seen to be negligible for the parameters in \fref{rho_evolution}, except for the case with a quadratic potential and $\Gamma_\phi/H_{\rm end}=10^{-5}$, where $\rho_A$ redshifting like radiation allows  $\rho_{A,{\rm pert}}$ to catch up to it before the end of reheating. This is consistent with what was seen in \fref{mechanisms} where for the $\phi^2$ potential perturbative decays cause our formalism to become unreliable after approximately 5 e-folds. In the quartic potential the perturbatively produced dark photons are safely negligible (since $\rho_A$ never enters a redshifting stage), however it is important to note that for smaller values of $\beta$ both contributions may be already comparable at the end of inflation. Because of this, even with a quartic potential it is important to assure that the value of $\beta$ is large enough for the hierarchy $\rho_A\gg\rho_{A,{\rm pert}}$ to be maintained.

After the end of reheating the inflaton behaves differently for the two potentials, as seen in the first row of~\fref{rho_evolution}. 
For $\phi^2$, the frequency $m_\phi\approx\sqrt{3}H_{end}$ is constant and larger than $\Gamma_\phi$, so the inflaton keeps oscillating with an exponentially decaying amplitude, with its kinetic and potential energies decreasing together. For $\phi^4$, instead, the frequency is proportional to the amplitude of the oscillations and decreases as the inflaton loses energy. 
In this run, reheating ends at $N\simeq 1.21$, and the inflaton crosses the minimum only once afterwards, at $N\simeq 1.28$. 
In the last fast swing its energy is mostly kinetic, and it is drained quickly (steep fall on the plot). The field climbs to its
last maximum at $N\simeq 1.65$, and the energy loss pauses there (small shoulder on the plot). It then rolls back
and picks up kinetic energy, which $\Gamma_\phi$ immediately drains. This is an overdamped regime with friction
balancing the force $\Gamma_\phi \dot\phi \simeq -\lambda_\phi \phi^3$. The late time solution is 
$\phi^2 \simeq \Gamma_\phi / (2\lambda_\phi t)$, so $\rho_\phi \propto V$ scales as $t^{-2}$. As we are already in
a radiation dominated universe with $t\propto a^2$, we have $\rho_\phi \propto a^{-4}$. 
The $a^{-4}$ behavior is a coincidence of overdamped $\phi^4$ plus radiation domination and is an artefact of a constant $\Gamma_\phi$.
In realistic models the decay rate would depend on the amplitude of the oscillations.
In both quadratic and quartic cases $\dot\phi$, and with it the instability parameter $\xi$, quickly becomes small, so dark photon production stops once the radiation-dominated era begins.

As we have observed in Figs.~\ref{fig:mechanisms} and \ref{fig:rho_evolution}, the homogeneous backreaction picture stops being a sufficient description of this dark matter model once the reheating phase reaches a certain duration. In the case of a quadratic potential this is due to perturbative decays from the inflaton into dark photons becoming relevant, and in the quartic potential it is due to the dark photon energy density becoming comparable with the total energy density of the universe and breaking the assumption of a fully homogeneous inflaton field, as well as the preheating of the inflaton's fluctuations caused by the inflaton self-interaction. Although we will continue to check for these conditions directly in the rest of this work, we have found that in the general case these phenomena start to take place before the end of reheating approximately when $\Gamma_\phi/H_{\rm end}\sim10^{-5}$. Despite this, it will also be necessary to keep track of the total amount of dark photons that are produced perturbatively and non-perturbatively, in order to maintain the hierarchy $\rho_A\gg\rho_{A,{\rm pert}}$.

To quantify the impact that the reheating phase really has on the dark matter parameter space, we parametrize how much $\rho_A$ changes after the end of inflation, by defining the parameter $\eps_A$ such that
\begin{equation}
\rho_A = \epsilon_A \rho_A^{(\rm end)},
\end{equation}
where $\epsilon_A^{(\rm RH)} = 1$ (that is, $\eps_A=1$ at the end of reheating) was assumed in previous studies \cite{Bastero-Gil:2018uel,Bastero-Gil:2021wsf}. In \fref{epsa} we show the results for $\epsilon_A$ as a function of the coupling $\beta$, evaluated at the beginning (black line) and the end (colored lines) of the reheating phase, comparing different values of the ratio $\Gamma_\phi/H_{\rm end}$. We take $\Gamma_\phi/H_{\rm end}=1$, which already yields a near-instantaneous reheating phase, as an upper limit in order to avoid pathological numerical behaviors that may arise when introducing such a large friction source in an infinitely short amount of time. The shape of these curves is independent of the value of $H_{\rm end}$, although the value of $\beta$ at which the system enters the strong backreaction regime is smaller for larger $H_{\rm end}$. For this reason we take $H_{\rm end}\simeq10^{9}\,\text{GeV}$ in this plot in order to show a wider range of values of $\beta$.

As a general rule, we see that $\epsilon_A$ grows with the coupling, which is to be expected since a larger coupling causes more dark photons to be produced during the reheating phase. We can see that already at the start of reheating the value of $\eps_A$ is different from a constant curve at $\epsilon_A=1$ since $\rho_A$ has already varied from its value at $\eps_H=1$ by the time the decay rate is introduced. This is simply a consequence of the fact that it is impossible to reheat the universe instantaneously and immediately as inflation ends using a perturbative decay rate. Although the exact shape of the $\Gamma_\phi\simeq H_{end}$ curve will depend on the condition chosen for the start of reheating (in our case $\eps_H=2$), any condition that waits until the inflaton has gained enough kinetic energy will always result in a curve different from a constant $\eps_A=1$.

We also observe a difference in behavior between the two types of potential, consistent with what was observed in Figs.~\ref{fig:mechanisms} and \ref{fig:rho_evolution}. In the $\phi^2$ case we see $\eps_A$ decreases as $\Gamma_\phi/H_{\rm end}$ becomes smaller, due to the fact that $\rho_A$ eventually stops being produced and is simply redshifted indefinitely until the end of reheating. Although increasing $\beta$ increases the value of $\eps_A$, for every value of $\beta$ there is always a certain $\Gamma_\phi$ below which $\eps_A<1$.

On the other hand, in the $\phi^4$ case, first we observe that in the instantaneous reheating limit $\eps_A$ is further away from a constant curve at $\eps_A=1$ than in the $\phi^2$ case due to reheating starting later, as can be seen in \fref{rho_evolution}. As opposed to the quadratic case, for a quartic potential there is a certain value of $\beta$ above which production always becomes larger as reheating efficiency is decreased, consistent with what was discussed above.

In general, we find that phenomena taking place between the end of inflation and the end of reheating can result in a change of up to 8 orders of magnitude in the value of $\rho_A$ at the end of reheating for a coupling within the range $\beta\leq9$ and a decay rate $\Gamma_\phi\geq 10^{-3} H_{\rm end}$.

\subsection{Energy density spectra}\label{sec:spec}

\begin{figure*}[t]
    \centering
    \includegraphics[width=0.7\textwidth]{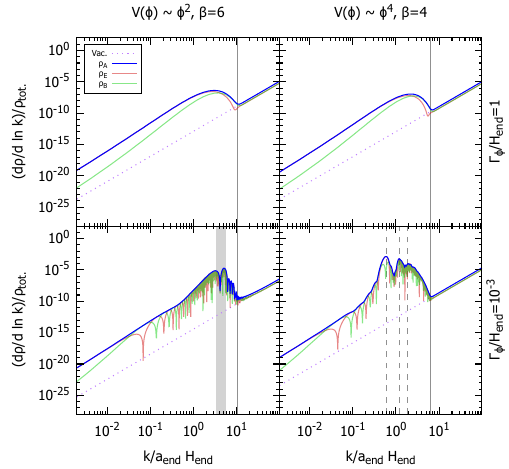}
    \caption{Electric, magnetic and total energy density spectra of the dark photon (including the contribution from both polarizations, $A_+$ and $A_-$), at the end of reheating for $H_{\rm end}\simeq10^{13}\,\text{GeV}$, comparing for different values of the decay rate $\Gamma_\phi$ as well as the different types of potential and values of the coupling $\beta$. The solid vertical line marks $k=k_h$. In the $\phi^4$ case, the dashed vertical lines mark the center of the first three parametric resonance bands, evaluated from \eqref{eq:res_phi4}. In the $\phi^2$ case we highlight in a gray band the range of momenta which get an enhancement from narrow resonance crossing, estimated from \eqref{eq:bandcrossing}.}
    \label{fig:spectra}
\end{figure*}

In this section we calculate the energy density spectra of the dark photon, separated into its electric $\rho_E=\frac{1}{2}\left\langle\boldsymbol{E}^2\right\rangle$ and magnetic $\rho_B=\frac{1}{2}\left\langle\boldsymbol{B}^2\right\rangle$ contributions,
\begin{equation}
    \frac{d\rho_A}{d\log k} = \frac{d\rho_E}{d\log k}+\frac{d\rho_B}{d\log k},
\end{equation}
 which can be obtained from Eqs.\eqref{eq:Esqrd} and \eqref{eq: Bsqrd},
\begin{align}
     \frac{d\rho_E}{d\log k}
    &=\frac{k^3}{4\pi^2a^2}\left(\left|\dot A_+\right|^2+\left|\dot A_-\right|^2\right),\\
    \frac{d\rho_B}{d\log k}&=\frac{k^5}{4\pi^2a^4}\left(\left|A_+\right|^2+\left|A_-\right|^2\right).
\end{align}
In \fref{spectra} we show the electric, magnetic, and total energy density spectra of the dark photon at the end of reheating. The spectra are normalized to the total energy density ($\rho_{tot}^{(\rm RH)} =3m_P^2H_{\rm RH}^2$) and shown as a function of the comoving momentum $k$ relative to the comoving Hubble horizon at the end of inflation, $1/a_{\rm end}H_{\rm end}$. We choose this normalization for continuity with previous works and easier comparison with previous works and because the range of relevant momenta for the spectra does not change after inflation ends.

The two plots in the top row of \fref{spectra} present the energy density spectra of the dark photon for a very fast (almost instantaneous) reheating period, $\Gamma_\phi\sim H_{\rm end}$, comparing the quartic and quadratic potential cases. These spectra are similar to those found in previous works at the end of inflation \cite{Bastero-Gil:2021wsf,Sobol:2021rlt}, showing a clear scale $k\sim a_{\rm end}H_{\rm end}$ around which the spectra peak and will contribute the most to the integrated energy density. Thus, in this case, we can assume that the energy density of the dark photon will be mostly populated with dark photons that possess a momentum of the order of the Hubble horizon at the end of inflation.

In the bottom row of \fref{spectra} we show the spectra for the same parameters, but with $\Gamma_\phi=10^{-3}H_{\rm end}$, which results in a longer reheating phase. Two kinds of features appear. At small $k$, the electric and magnetic spectra display a regular comb of dips, evenly spaced in $k$ and in antiphase, while their sum is unaffected. This comb is not produced directly by the inflaton oscillations. At the end of inflation, all the tachyonically amplified modes share a common phase, with $|\dot A|\gg k|A|/a$, i.e.\ they are dominated by the electric field. For the modes with small $k$, the force exerted by the oscillating inflaton in Eq.~\eqref{eq:TransvModes} reverses its sign every half-oscillation, faster than the modes can respond, i.e.\ the half-period is short compared with $\big(k\,\beta\,|\phi'|_{\rm max}/m_P\big)^{-1/2}$. Its effect therefore averages out over each period, and these modes evolve essentially freely during reheating. At the end of reheating this gives $d\rho_B/d\ln k\propto\sin^2(k\Delta\tau)$ and $d\rho_E/d\ln k\propto\cos^2(k\Delta\tau)$, where $\Delta\tau=\int dN/(aH)$ is the conformal duration of reheating. Only the modes with $k\Delta\tau\gtrsim1$, i.e.\ those that are inside the horizon at the end of reheating, $k\gtrsim a_{\rm RH}H_{\rm RH}$, have had time to oscillate, and the comb starts there. For the quadratic potential, the spacing of the dips, $\pi/\Delta\tau$, indeed agrees with the conformal duration of reheating. The comb keeps evolving after reheating and leaves no imprint on the total spectrum.

At larger $k$, instead, the total spectrum itself loses its single-peak structure, as a result of the parametric resonance. 
For $\phi^4$, the additional peaks at $k\simeq0.6$ and $1.2\,a_{\rm end}H_{\rm end}$ coincide with the first two resonance bands, Eq.~\eqref{eq:res_phi4}, which are fixed in comoving momentum. For $\phi^2$ the bands sweep through the spectrum, and the features are spread over the range of band crossings, Eq.~\eqref{eq:bandcrossing}, which contains the two maxima at $k\simeq3.3$ and $5.2\,a_{\rm end}H_{\rm end}$. Above $k_h$ the bands act only on modes that are essentially in the vacuum, and the spectrum is unaffected.

The same resonance also amplifies the subdominant polarization, as shown in \fref{spectraAmin}. Unlike $A_+$, the mode $A_-$ was never tachyonically amplified during inflation, so all the structure in its spectrum is generated during reheating. Its peaks sit at the same momenta as those of the total spectrum. This confirms that the latter are produced by the resonance, which amplifies both polarizations at the same rate. Since $A_-$ starts from a much smaller amplitude, it remains subdominant by a few orders of magnitude, and the chirality generated at the end of inflation is preserved.

\begin{figure*}[t]
    \centering
    \includegraphics[width=0.7\textwidth]{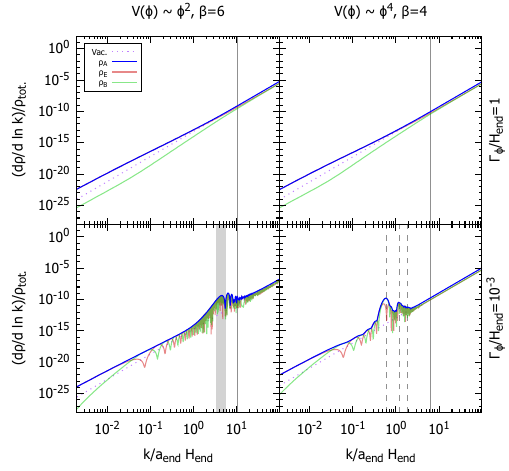}
    \caption{Same as \fref{spectra}, but for the contribution
    of the subdominant polarization $A_-$ only.}
    \label{fig:spectraAmin}
\end{figure*}

\section{Dark matter parameter space}\label{sec:pspace}

After the end of the reheating phase, the universe goes on to become radiation-dominated and begins its known thermal history until present day. In order to study the evolution of the different components of the universe, we use the relation between the radiation energy density $\rho_R$ and the temperature $T$ of the thermal bath:
\begin{equation}
    \rho_R=g_*\frac{\pi^2}{30}T^4,
\end{equation}
where $g_*$ is the effective number of radiation degrees of freedom, which can be assumed to be that of the Standard Model. We take the value $g_*=106.75$ which can be used as long as $T>173.3\,\text{GeV}$ \cite{DoF}. Using this relation, we obtain the reheating temperature $T_{\rm RH}$ from $\rho_R$ at the end of the reheating phase.

By assuming that the dark photons account for the full relic abundance of dark matter at present time, we can find an implicit relation between the two free parameters in our dark matter model, the coupling $\beta$ and the mass $m_A$. First, we track the evolution of the dark photon energy density up until present day \cite{Bastero-Gil:2021wsf},
\begin{equation}
    \rho_A(T)=\begin{cases}
        \rho_A (T_{\rm RH})\left(\frac{T}{T_{\rm RH}}\right)^4 & T>\overline T\\
        \rho_A (T_{0}) \left(\frac{T}{T_{0}}\right)^3 & T<\overline T ,
    \end{cases}\label{eq:photonredshift}
\end{equation}
where $T_0$ is the present day temperature of the universe, and $\overline T$ is the temperature at which the dark photons become non-relativistic. This will occur when their physical momentum $q=k/a$ is equal to their mass,
\begin{equation}
	q (\overline T) =q(T_{\rm RH})\frac{\overline T}{T_{\rm RH}}=m_A\Rightarrow\overline T=m_A\frac{T_{\rm RH}}{q_{\rm RH}},\label{eq:Tbar}
\end{equation}
which takes place at different times (and temperatures) for different values of the comoving momentum $k$.

As seen in the previous section, in scenarios with fast and highly efficient reheating most of the contribution to the energy density will become non-relativistic around the same time as the mode at the peak of the spectrum. However, oscillations taking place during a less efficient reheating phase can cause the spectrum to stop having a clear peak. Therefore, we will take the energy density spectrum as a distribution and compute the average value of $q_{\rm RH}$ using the integral,
\begin{equation}
	\left< q\right>
	=
	\frac{1}{\rho_A}
	\intop d\ln k
	\frac{k}{a}
	\left(
	\frac{d\rho_E}{d\ln k}
	+
    \frac{d\rho_B}{d\ln k}
	\right),
\end{equation}
and evaluate it at the end of reheating in order to find $\left\langle q_{\rm RH}\right\rangle\equiv\left\langle q(T_{\rm RH})\right\rangle$. By equating the two cases in (\ref{eq:photonredshift}) at $T=\overline T$ and combining with equation (\ref{eq:Tbar}) we find
\begin{equation}
    m_A=\left< q_{\rm RH}\right>\frac{\rho_\text{CDM}}{\rho_A(T_{\rm RH})}\frac{\Omega_A}{\Omega_\text{CDM}}\left(\frac{T_{\rm RH}}{T_0}\right)^3,
\end{equation}
where we have expressed $\rho_A(T_0)$ in terms of the present-day energy density of cold dark matter $\rho_\text{CDM}$ using the corresponding density parameters, defined as $\Omega_X=\rho_X/3 H_0^2m_P^2$. Taking the measured values of the radiation temperature today $T_0=2.35\cdot 10^{-13}\,\text{GeV}$ \cite{Fixsen:2009ug} and observed DM relic abundance $\rho_\text{CDM}=9.4\cdot 10^{-48}\,\text{GeV}^4$ \cite{Planck:2018vyg} gives,
\begin{equation}
    m_A=7.22 \cdot 10^{-10} \,\text{GeV}\,\frac{\left< q_{\rm RH}\right>T_{\rm RH}^3}{\rho_A(T_{\rm RH})}\frac{\Omega_A}{\Omega_\text{CDM}},\label{eq:mA}
\end{equation}
with the ratio $\Omega_A/\Omega_\text{CDM}$ signifying the fraction of DM that is constituted by dark photons. We will assume $\Omega_A=\Omega_\text{CDM}$ from this point on, so that imposing this condition on the dark photon mass ensures that its total energy density will reproduce the necessary relic abundance at present day. Since $k_h$ does not grow during reheating, the position of the spectrum does not shift during this phase either, meaning that $\left\langle q_{\rm RH}\right\rangle$ will redshift approximately like
\begin{equation}
    \left\langle q_{\rm RH}\right\rangle
    \sim
    \left\langle q_{\rm end}\right\rangle
    e^{-N_{\rm RH}},
\end{equation}
with $N_{\rm RH}$ being the number of e-folds from the end of inflation to the end of reheating. Given this behavior, we can see in Eq.~\eqref{eq:mA} that a longer, less efficient reheating period (that is, larger $N_{\rm RH}$ and smaller $T_{\rm RH}$) results in a smaller dark photon mass, but so does a larger value of its energy density at the end of reheating. 

When exploring the resulting dark matter parameter space one must ensure that the dark vector becomes non-relativistic after reheating has ended, but before matter-radiation equality, therefore keeping the hierarchy $T_{\rm RH}>\overline T>T_{eq}$. We can express this condition as an allowed range for the dark photon mass,
\begin{equation}
\left\langle q_{\rm RH}\right\rangle 
\frac{T_{eq}}{T_{\rm RH}}
< m_A < 
\left\langle q_{\rm RH} \right\rangle ,\label{eq:TempConstraint}
\end{equation}
where $T_{eq}=7.96\cdot 10^{-10}\,\text{GeV}$ \cite{Dupac:2004pg,Planck:2018vyg}. Another model-independent bound on the Hubble parameter at the end of inflation can be derived from observational bounds on 
the tensor-to-scalar ratio, $r<0.036$ \cite{Planck:2018vyg,BICEP:2021xfz} effectively resulting in $H_{\rm end}<H_*\lesssim 4.7\cdot 10^{13}\, \text{GeV}$.  Additionally, one must remember that due to the homogeneous backreaction picture, our description relies on several assumptions mentioned in previous sections: that the hierarchies $\rho_\phi\gg\rho_A$ and $\Gamma_\phi\gg\Gamma_A$, where $\Gamma_A$ is defined in \eqref{eq:GammaA}, are maintained at all times, and that the coupling $\beta$ is kept under the bound $\beta < 12$. 
Exploring parameter space beyond these limits requires lattice calculations~\footnote{Recent lattice codes utilized previously in the context of magnetogenesis can incorporate a U(1) gauge field with an axion like coupling to the inflaton such as CosmoLattice 2.0 \cite{Baeza-Ballesteros:2026uao} and PencilCode \cite{PencilCode:2020eyn} which also includes the Schwinger effect \cite{Iarygina:2025ncl} and a massive gauge boson during inflation \cite{Ferreira:2026klg}.}.

\begin{figure*}[t]
    \centering
\includegraphics[width = 0.6\textwidth]{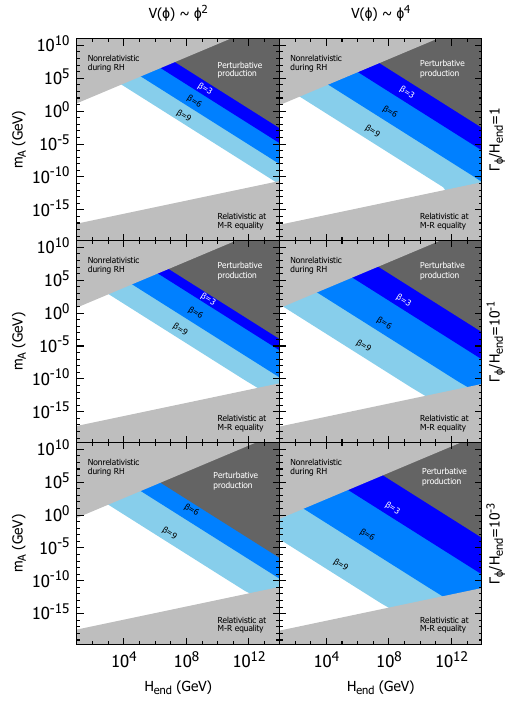}
    \caption{Contour plots of the relic abundance as a function of the dark photon mass $m_A$ and the Hubble parameter at the end of inflation, $H_{\rm end}$, as obtained from Eq.~\eqref{eq:mA} when assuming $\Omega_A=\Omega_\text{CDM}$. Lighter gray areas mark the regions ruled out by \eqref{eq:TempConstraint}, while darker gray areas are regions where the perturbatively produced contribution to the dark photons is no longer negligible. We compare numerical results for different values of the coupling $\beta$ and of the reheating decay rate $\Gamma_\phi$ as well as for the case of a quadratic and a quartic potential.}
    \label{fig:parameter_space}
\end{figure*}
In \fref{parameter_space} we present the available dark photon dark matter parameter space in the form of contours of Eq.~\eqref{eq:mA} (under the assumption that $\Omega_A=\Omega_\text{CDM}$) as a function of $m_A$ and the Hubble parameter at the end of inflation. We compare these results, which were obtained numerically, for different values of the coupling $\beta$, the decay rate $\Gamma_\phi$, and the cases of a quadratic and quartic potential. We also show the regions ruled out by the constraints in Eq.~\eqref{eq:TempConstraint} as light gray areas in each case, as well as regions where perturbatively produced dark photons have a non-negligible contribution to the energy density, marked in dark gray. Although we include an analytical estimation for the latter in the appendix (Eq.~\eqref{eq:ratioquad} for a quadratic potential, and Eq.~\eqref{eq:ratioquart} for a quartic potential), their position in this plot has been estimated numerically.

The first row shows the results for $\Gamma_\phi=H_{\rm end}$, which corresponds to near-instantaneous reheating. In this scenario, reheating ends very shortly after it starts at $\epsilon_H=2$, before the inflaton reaches the minimum of its potential and begins to oscillate. This means that the only difference between the dark photon energy density at the end of inflation and at the end of reheating is due to the small intermediate phase between $\eps_H=1$ and $\eps_H=2$. Depending on the coupling this gives a difference of up to one or two orders of magnitude, as can be seen in \fref{epsa}. Thus these plots reproduce, up to this difference, previous analytical results assuming instantaneous reheating at the end of inflation and a very high reheating efficiency \cite{Bastero-Gil:2021wsf}.

As reheating efficiency is reduced, we can again see a difference in behavior with respect to the type of potential. As was shown in the previous section, in the $\phi^4$ case dark photons are continuously produced as long as reheating continues. Thus, delaying the end of reheating results in a smaller value of $T_{\rm RH}$ while keeping $\rho_A^{(\rm RH)}$ approximately constant. This way, we see that in the case of a quartic potential reducing reheating efficiency simply results in smaller values of the dark matter mass. In the case of a quadratic potential, reducing $\Gamma_\phi$ also results in smaller masses at first, since the dark photons continue to be produced during the first oscillations of reheating. However, since after a certain point the dark photons begin to redshift, we see that further decreasing $\Gamma_\phi$ does not necessarily lead to even smaller masses and in fact eventually causes the masses to increase again. This can be seen in the $\beta=6$ line in the $\Gamma_\phi/H_{\rm end}=10^{-3}$ case, which shows masses one order of magnitude above those in the $\Gamma_\phi/H_{\rm end}=10^{-1}$ case.
    
Given these results, we arrive at a final mass range of $10^{-14}\,\text{GeV}\lesssim m_A\lesssim 10^{7}\,\text{GeV}$ spanning over 20 orders of magnitude. This is a similar range to the one found in previous studies \cite{Bastero-Gil:2021wsf} which assumed instantaneous reheating. However, as we have shown here, the lowest end of the mass range is only achievable when there is a long oscillatory period where the dark photons are continuously produced which only seems to be possible in the quartic potential case. Thus the dark photon mass itself contains information about reheating and, combined with measurements of the energy density spectra, would potentially allow us to infer the inflationary potential.

The potential presence of a kinetic mixing between the dark photon and the SM photon opens up direct detection possibilities, but it must be small enough for the dark photon not to thermalize with the SM. Furthermore, decays to positron-electron pairs open up if its mass is above $1\,\text{MeV}$ unless the kinetic mixing is far too small to be detected. On the other hand, a mass below the $1\,\text{MeV}$ threshold results in long-lived dark photons even with the presence of a kinetic mixing allowing them to be potentially detectable without spoiling the dark matter model \cite{Caputo:2026pdw,Baudis:2025yva}. As seen in Fig. 15 of \cite{Caputo:2026pdw}, our results are compatible with a potentially detectable window around masses of $10^{-7}\,\text{eV}\lesssim m_A\lesssim 100\,\text{eV}$ and a kinetic mixing $\varepsilon\lesssim10^{-10}$, with the latter being small enough to avoid thermalization with the SM. A detailed study of direct detection prospects is also left to future work. 

\section{Summary and Discussion} \label{sec:sum} 

In this work we have used numerical simulations in order to study the impact of the reheating phase on the inflationary vector dark matter production mechanism introduced in \cite{Bastero-Gil:2018uel} which assumed instantaneous reheating. In particular, we have tracked the evolution of the energy density of a (massive) $U(1)$ gauge boson, produced via a tachyonic instability at the end of inflation, through reheating while the inflaton oscillates about its minimum and into a radiation dominated era until today. Limiting ourselves to the homogeneous backreaction picture, we have used the gradient expansion formalism together with a mode-by-mode integration in order to obtain a description of the system's evolution from CMB scales to the end of the reheating phase. We find the energy density at the end of reheating can be increased or decreased by several orders of magnitude with respect to the assumption of instantaneous reheating and find a final vector dark matter mass range of $10^{-14}\,\text{GeV}\lesssim m_A\lesssim 10^{7}\,\text{GeV}$ (see \fref{parameter_space}).

We find a notable difference in the behavior of the dark photon energy density during reheating depending on whether the inflaton potential can be approximated by a quadratic ($\phi^2$) or quartic ($\phi^4$) monomial during its oscillatory phase (see \fref{rho_evolution}). In the quadratic case, dark photon production continues during the first couple of e-folds of the oscillatory phase, but then quickly diminishes causing the dark photons to start redshifting until the end of reheating. In the quartic case, however, dark photon production continues uninterrupted through the end of reheating until its energy density eventually surpasses the inflaton's and breaks the homogeneous backreaction assumption. This enhanced production in the quartic case leads to smaller possible masses for vector dark matter than in the quadratic case. 

We also find distinct features in the dark matter power spectrum caused by the inflaton's oscillations during reheating, due to the different behavior of the 
parametric resonance in the two scalar potentials
(see \fref{spectra}).
For the quadratic case, the location and width of the resonance bands change over time, sweeping quickly through all the comoving spectrum. This led to limited features around and above the specturm's peak defined at the end of inflation. In contrast, for the quartic case the resonance bands are fixed in specific comoving momentum. Since these momenta are amplified at a constant rate per e-fold as long as the inflaton oscillates, the spectrum obtained growing peaks at those momenta for long reheating periods.

Despite the differences between the inflationary potentials on the dark vector spectrum at the end of reheating, they share the common trait of the dark vector abundance being predominately populated by a definite helicity, in the region of parameters explored.
A dedicated analysis of post-reheating dynamics, and in particular of the structure formation epoch, is important to understand if this disparity between helicities survives through the cosmic history. We leave this to future work.

The results obtained in this study are constrained by several assumptions, such as a homogeneous inflaton field at all times, or a purely perturbative reheating process. Relaxing these assumptions could potentially increase the available parameter space and in particular allow for much lighter vector dark matter masses. This requires going into the strong backreaction regime where lattice computations are needed and is left to ongoing work \footnote{A newer version of the gradient expansion including inflaton perturbations presented in a recent study also shows a promising formalism capable of capturing the fragmentation of the inflaton field \cite{Domcke:2026fjo}.}. Nevertheless, the results of the current study make it clear that the duration of reheating and the inflationary potential play a key role on the parameter space of this vector dark matter production mechanism as well as on the shape of the resulting dark matter power spectrum. 

\vspace{0.2cm}
\emph{Note added:} During the preparation of this work, we became aware of \cite{Gustavo} which also considers oscillation effects of the inflaton during reheating on vector dark matter.
Their work focuses on $\alpha$-attractor models which have a much steeper potential than those considered here leading to quantitatively different predictions and complementary results. 
We would like to thank the authors of \cite{Gustavo} for interesting discussions, for sharing their preliminary draft,
and for coordinating the joint arXiv submission of our
studies.

The effect of the inflaton oscillations on dark photons with a Chern-Simons coupling to the inflaton was also studied recently in \cite{Jenks:2026ztv}, in the context of cosmological gravitational particle production of massive dark photons. That work is complementary to ours. It considers much weaker couplings, for which tachyonic amplification during inflation enhances the dominant helicity by at most a couple of orders of magnitude, and computes the number of produced particles mode by mode, including the longitudinal polarization, on a fixed background without backreaction. It considers a quadratic potential only, and models reheating as an abrupt transition to radiation domination. Here, instead, we focus on couplings for which tachyonic production at the end of inflation is exponentially large, include its backreaction on the inflaton in the homogeneous approximation, model reheating through a decay rate into radiation, and consider both quadratic and quartic potentials.

\vspace{0.3cm}
\noindent
{\bf Acknowledgments:}~The authors thank Gustavo Henrique dos Santos and Chee Sheng Fong, whose work overlaps with ours, for insightful discussion, as well as António Torres Manso, Manel Masip and Jose Santiago for useful comments and discussions. This work has been partially supported by FEDER/UE (grants PID2022-139466NB-C21/C22 and PID2025-170216NB-I00) (R.V.M.) as well as by Grant PID2025-PID2025-168050NB-I00 funded by MICIU/AEI/10.13039/501100011033 (M.B.G.), FCT-CERN grant No. CERN/FIS-PAR/0027/2021 (M.B.G., P.F.), FCT Grant No. SFRH/BD/151475/2021 (P.F.). The work of L.U. is supported by the Slovenian Research Agency under the research core funding No. P1-0035 and the research grant J1-60026, and the work of P.G.O. is supported by contract number 10025 under the 2026 Programa de Empleo Juvenil of the University of Granada. This article is based upon work from COST Action COSMIC WISPers CA21106, supported by COST (European Cooperation in Science and Technology). R.V.M. also thanks the TD-Lee institute in Shanghai for their hospitality and partial support as well as participants of the NEXUS 2026 workshop for useful discussions and stimulating atmosphere while part of this work was completed.

\appendix

\section{Analytical estimation of the perturbative production of dark photons}

In this appendix we include an analytical estimation for the fraction of dark photons that are produced via direct particle decays from the inflaton; 
in particular we study whether the perturbative contribution to the dark photons energy density $\rho_{A,\rm pert}$, coming from the inflaton decay into two dark photons, may overcome that of the non-perturbative $\rho_A$,  which we have studied in detail in this work. Reheating is controlled by the decay rate of the inflaton field $\Gamma_\phi$ into radiation, so we first check that indeed $\rho_{A,\rm pert} \ll \rho_{R}$. 

We treat the contribution $\rho_{A,\rm pert} $ as a separate component. The evolution Eqs. for the energy densities of the inflaton $\rho_\phi$, radiation $\rho_R$ and perturbative dark photons $\rho_{A,\rm pert} $ during reheating are given by:
\begin{eqnarray}
\dot \rho_\phi + 3 H (\rho_\phi + p_\phi) &\simeq& - (\Gamma_\phi + \Gamma_A) (\rho_\phi + p_\phi) \,, \\
\dot \rho_R + 4 H \rho_R &\simeq&   \Gamma_\phi (\rho_\phi + p_\phi) \,, \\
\dot \rho_{A,\rm pert} + 4 H \rho_{A,\rm pert} &\simeq&   \Gamma_A (\rho_\phi + p_\phi) \,, 
\end{eqnarray}
where $\Gamma_A$ is given by Eq. \eqref{eq:GammaA}, and it would be anyway much smaller than $\Gamma_\phi$, the main decay rate controlling the duration of reheating. With respect to Eqs.~\eqref{eq:inflatondensity} and \eqref{eq:rhoA} we have averaged the value $\langle \dot \phi^2 \rangle = \rho_\phi + p_\phi$, and neglected the subdominant contribution proportional to $\langle E\cdot B \rangle$ in the inflaton's equation of motion. For the analytical estimation, both $\Gamma_\phi$ and $\Gamma_A$ are switched on at $\epsilon_H=2$ after inflation ends. 

Given a quadratic potential, during the oscillations the inflaton's energy density behaves like matter ($p_\phi=0$, $\rho_\phi \propto a^{-3}$), $\Gamma_A$ and $\Gamma_\phi$ are constants, and  $\rho_R$ and $\rho_{A,\rm pert}$ are then both redshifted as $a^{-3/2}$ during reheating, therefore their ratio is given by that of the decay rates:
\begin{equation}
\frac{\rho_{A,\rm pert}}{\rho_R}=\frac{\Gamma_A}{\Gamma_\phi}=   
1.56 \cdot 10^{-11} \left(\frac{\beta}{6}\right)^2 \left( \frac{H_{\rm end}}{10^{13}\,\gev}\right)^2 \left(\frac{H_{\rm end}}{\Gamma_\phi}\right) \,, \label{RhoRA_quad}
\end{equation}
where we have used the condition $\epsilon_\phi=m_P^2(\partial_\phi V/V)^2/2\simeq 1$ at the end of inflation to replace $m_\phi \simeq \sqrt{3}H_{\rm end}$ in $\Gamma_A$. Thus, unless $\Gamma_\phi/H_{\rm end} \ll 10^{-11}$ we find, as expected, that $\rho_{A,\rm pert} \ll \rho_R$.

For the quartic potential, the mass squared of the inflaton field is given by  the second derivative of the potential, $m_\phi^2 =3 \lambda_\phi \phi^2$. The inflaton energy density behaves now as radiation during reheating, such that $\phi \propto a^{-1}$,  $m_\phi \propto a^{-1}$ and $\Gamma_A \propto a^{-3}$. The energy densities $\rho_R$ and $\rho_{R,\rm pert}$ will have different behaviors during the oscillations, $\rho_{R}\propto a^{-2}$, $\rho_{A,\rm pert}\propto a^{-4}$, and the ratio $\rho_{A,\rm pert}/\rho_R$  will decrease during reheating from its initial value:
\begin{equation}
\frac{\rho_{A,\rm pert}}{\rho_{R}} = \frac{2\Gamma_{A,0}}{\Gamma_{\phi}} \simeq 4\cdot 10^{-9} \left(\frac{\beta}{6}\right)^2 \left(\frac{H_{\rm end}}{10^{13} \,\gev} \right)^2 \left(\frac{H_{\rm end}}{\Gamma_\phi}\right) \,. 
\end{equation}
where the subscript ``0'' denotes the start of reheating at $\epsilon_H=2$. Again, unless $\Gamma_\phi/H_{\rm end}< 10^{-9}$ we always have $\rho_{A,\rm pert} \ll \rho_{R}$.
Therefore, for both quadratic and quartic potentials, we can safely neglect $\Gamma_A$ in the equation of motion of the inflaton condensate and $\rho_{A,\rm pert}$ in the total energy density when modeling reheating. 

What we want to compare is the relative contribution of $\rho_{A,\rm pert}$ to the non-perturbative one $\rho_A$, which has been computed numerically through this work. In particular we will use the ratio $\epsilon_A^{(\rm RH)} = \rho_A^{(\rm RH)}/\rho_A^{\rm end}$  shown in Fig. \eqref{fig:epsa}, which depends on $\beta$ and $\Gamma_\phi/H_{\rm end}$ but it is independent of $H_{\rm end}$. Then the ratio between the dark photons' non-perturbative/perturbative contributions by the end of reheating can be written as:
\begin{equation}
\frac{\rho_A^{(\rm RH)}}{\rho_{A,\rm pert}^{(\rm RH)}} = \left(\epsilon_A \cdot \frac{\rho_A^{\rm end}}{\rho_\phi^{\rm end}}\right) \left(\frac{H_{\rm end}}{H_0} \right)^2 \left(\frac{\rho_{\phi,0}}{\rho_{A,\rm pert}^{(\rm RH)}} \right)\,. \label{eq:ratio}
\end{equation}
For the quadratic potential with constant $\Gamma_A$ we have:
\begin{equation}
\frac{\rho_{\phi,0}}{\rho_{A,\rm pert} ^{(\rm RH)}} = \frac{25 H_0^2}{4 \Gamma_\phi \Gamma_A} \,,
\end{equation}
and plugging this and the expression for $\Gamma_A$ into Eq. \eqref{eq:ratio} we have:
\begin{equation}
\frac{\rho_A^{(\rm RH)}}{\rho_{A,\rm pert} ^{(\rm RH)}} \simeq 2.24\left(\epsilon_A^{(\rm RH)} \cdot \frac{\rho_A^{\rm end}}{H_{\rm end}^4}\right) \left(\frac{6}{\beta}\right)^2 \left(\frac{H_{\rm end}}{\Gamma_\phi} \right) \,. \label{eq:ratioquad}
\end{equation}
Notice that $\rho_A^{\rm end} \propto H_{\rm end}^4$, and therefore the factor ($\epsilon_A \rho_A^{\rm end}/H_{\rm end}^4$) is independent of $H_{\rm end}$. Its dependence on the parameters $\Gamma_\phi/H_{\rm end}$ and $\beta$ can be easily extracted from our numerical simulations. For $\beta \gtrsim 2$ it can be well fitted as an exponential:
\begin{equation}
\epsilon_A^{(\rm RH)} \cdot \frac{\rho_A^{\rm end}}{H_{\rm end}^4} \simeq A_0 {\rm exp}(A_1 \beta) \,,
\end{equation}
where the coefficients $A_0$ and $A_1$ are just functions of the ratio $\Gamma_\phi/H_{\rm end}$, such that $\rho_A^{(\rm RH)}/\rho_{A,\rm pert} ^{(\rm RH)}$ grows with $\Gamma_\phi$. Therefore, imposing that Eq. \eqref{eq:ratioquad} be larger than one we can derive a lower bound on $\Gamma_\phi/H_{\rm end}$ depending on the value of $\beta$. 
Numerically we obtain:
\begin{equation}
\left(\frac{\Gamma_\phi}{H_{\rm end}}\right)^{min}_{quad} \simeq 1.5\, {\rm exp}(-1.9 \beta) \,, \label{eq:limit_quadratic}
\end{equation}
which for example for $\beta=3$ gives the constraint $\Gamma/H_{\rm end} \gtrsim 5\cdot 10^{-3}$, while for $\beta=6$ we have $\Gamma/H_{\rm end} \gtrsim 2\cdot 10^{-5}$. This reproduces well the trend numerically obtained in Fig. \eqref{fig:parameter_space} for the quadratic oscillations: the longer reheating, the stronger has to be non-perturbative dark photon production during inflation, i.e., the larger the value of $\beta$. 

Repeating the same exercise for the quartic potential we get: 
\begin{equation}
\frac{\rho_A^{(\rm RH)}}{\rho_{A,\rm pert} ^{(\rm RH)}} \simeq 0.88 \left(\epsilon_A^{(\rm RH)} \cdot \frac{\rho_A^{\rm end}}{H_{\rm end}^4}\right) \left(\frac{6}{\beta}\right)^2  \left(\frac{H_{\rm end}}{\Gamma_\phi} \right)^2 \,, \label{eq:ratioquart}
\end{equation}
with some dependence on the variation of the Hubble parameter and the field value between the end of inflation and the onset of the oscillations, $H_0/H_{\rm end} \simeq 0.394$, $\phi_{\rm end}/\phi_0 \simeq 1.895$. The factor ($\epsilon_A \rho_A^{\rm end}/H_{\rm end}^4$) grows like $(\Gamma_\phi/H_{\rm end})^\alpha$, but with  $\alpha \lesssim 2$. Therefore, instead of a lower bound on the value of $\Gamma_\phi$, we obtain an upper bound, given approximately by:
\begin{equation}
\left(\frac{\Gamma_\phi}{H_{\rm end}}\right)^{max}_{quart} \simeq 0.1\, {\rm exp}(1.2 \beta) \,,
\end{equation}
and for example for $\beta=2$ we have $\Gamma_\phi/H_{\rm end} \simeq O(1)$.

\bibliographystyle{apsrev}
\bibliography{NoSchwingerReheatingRefs}

@unpublished{Gustavo,
    author = {Henrique dos Santos, Gustavo and Sheng Fong, Chee and Hossain Rahat, Moinul},
    title = {Post-inflationary enhancement of vector dark matter production},
    year = {2026},
    note = {Work in preparation}
}

@article{Jenks:2026ztv,
    author = "Jenks, Leah and Kamionkowski, Marc and Kolb, Edward W.",
    title = "{Resonant Gravitational Production of Chiral Dark Photons}",
    eprint = "2609.22423",
    archivePrefix = "arXiv",
    primaryClass = "hep-ph",
    month = "9",
    year = "2026"
}

@article{Cline:2024wja,
    author = "Cline, James M. and Herrera, Gonzalo",
    title = "{Plausible constraints and inflationary production for dark photons}",
    eprint = "2409.13818",
    archivePrefix = "arXiv",
    primaryClass = "hep-ph",
    reportNumber = "CERN-TH-2024-154",
    doi = "10.1103/PhysRevD.112.035023",
    journal = "Phys. Rev. D",
    volume = "112",
    number = "3",
    pages = "035023",
    year = "2025"
}

@article{Cline:2026eps,
    author = "Cline, James M. and Herrera, Gonzalo and Roux, Jean-Samuel",
    title = "{Dark Photons from Perturbative Decay of a Misaligned Higgs Field}",
    eprint = "2607.29682",
    archivePrefix = "arXiv",
    primaryClass = "hep-ph",
    reportNumber = "FERMILAB-PUB-26-0618-V",
    month = "7",
    year = "2026"
}

@article{Arza:2026rsl,
    author = "Arza, A. and others",
    title = "{The COSMIC WISPers White Paper: The physics case for Weakly Interacting Slim Particles}",
    eprint = "2603.03433",
    archivePrefix = "arXiv",
    primaryClass = "hep-ph",
    reportNumber = "BARI-TH/784-26, CERN-TH-2026-016, IPPP/26/13, IFT-UAM/CSIC-26-13, KCL-PH-TH/2026-04, KEK-Cosmo-0411, KEK-TH-2804, LAPTH-008/26, MPP-2026-21, RESCEU-5/26, SLAC-PUB-260219, ST/T006994/1, ST/Y004531/1",
    month = "3",
    year = "2026"
}

@article{Greene:1997fu,
    author = "Greene, Patrick B. and Kofman, Lev and Linde, Andrei D. and Starobinsky, Alexei A.",
    title = "{Structure of resonance in preheating after inflation}",
    eprint = "hep-ph/9705347",
    archivePrefix = "arXiv",
    reportNumber = "SU-ITP-97-19, IFA-97-29",
    doi = "10.1103/PhysRevD.56.6175",
    journal = "Phys. Rev. D",
    volume = "56",
    pages = "6175--6192",
    year = "1997"
}

@article{Bastero-Gil:2018uel,
      author         = "Bastero-Gil, Mar and Santiago, Jose and Ubaldi, Lorenzo
                        and Vega-Morales, Roberto",
      title          = "{Vector dark matter production at the end of inflation}",
      journal        = "JCAP",
      volume         = "1904",
      year           = "2019",
      number         = "04",
      pages          = "015",
      doi            = "10.1088/1475-7516/2019/04/015",
      eprint         = "1810.07208",
      archivePrefix  = "arXiv",
      primaryClass   = "hep-ph",
      reportNumber   = "UG-FT 328-18, CAFPE 198-18, SISSA 44/2018/FISI",
      SLACcitation   = "%%CITATION = ARXIV:1810.07208;%%"
}

@article{Bastero-Gil:2021wsf,
    author = "Bastero-Gil, Mar and Santiago, Jose and Vega-Morales, Roberto and Ubaldi, Lorenzo",
    title = "{Dark photon dark matter from a rolling inflaton}",
    eprint = "2103.12145",
    archivePrefix = "arXiv",
    primaryClass = "hep-ph",
    reportNumber = "UG-FT 333-21, CAFPE 203-21",
    doi = "10.1088/1475-7516/2022/02/015",
    journal = "JCAP",
    volume = "02",
    number = "02",
    pages = "015",    year = "2022"
}

@article{Bastero-Gil:2023htv,
    author = "Bastero-Gil, Mar and Ferraz, Paulo B. and Ubaldi, Lorenzo and Vega-Morales, Roberto",
    title = "{Super heavy dark matter from inflationary Schwinger production}",
    eprint = "2311.09475",
    archivePrefix = "arXiv",
    primaryClass = "hep-ph",
    reportNumber = "UG-FT 328-23, CAFPE 198-23, CA21106",
    doi = "10.1103/PhysRevD.110.095019",
    journal = "Phys. Rev. D",
    volume = "110",
    number = "9",
    pages = "095019",
    year = "2024"
}

@article{Bastero-Gil:2023mxm,
    author = "Bastero-Gil, Mar and Ferraz, Paulo B. and Ubaldi, Lorenzo and Vega-Morales, Roberto",
    title = "{Schwinger dark matter production}",
    eprint = "2312.15137",
    archivePrefix = "arXiv",
    primaryClass = "hep-ph",
    reportNumber = "UG-FT 329-23, CAFPE 199-23, SISSA 44/2020/FISI, CA21106",
    doi = "10.1088/1475-7516/2024/10/078",
    journal = "JCAP",
    volume = "10",
    pages = "078",
    year = "2024"
}

@article{BICEP:2021xfz,
    author = "Ade, P. A. R. and others",
    collaboration = "BICEP, Keck",
    title = "{Improved Constraints on Primordial Gravitational Waves using Planck, WMAP, and BICEP/Keck Observations through the 2018 Observing Season}",
    eprint = "2110.00483",
    archivePrefix = "arXiv",
    primaryClass = "astro-ph.CO",
    doi = "10.1103/PhysRevLett.127.151301",
    journal = "Phys. Rev. Lett.",
    volume = "127",
    number = "15",
    pages = "151301",
    year = "2021"
}

@article{Nelson:2011sf,
      author         = "Nelson, Ann E. and Scholtz, Jakub",
      title          = "{Dark Light, Dark Matter and the Misalignment Mechanism}",
      journal        = "Phys. Rev.",
      volume         = "D84",
      year           = "2011",
      pages          = "103501",
      doi            = "10.1103/PhysRevD.84.103501",
      eprint         = "1105.2812",
      archivePrefix  = "arXiv",
      primaryClass   = "hep-ph",
      SLACcitation   = "%%CITATION = ARXIV:1105.2812;%%"
}

@article{Arias:2012az,
      author         = "Arias, Paola and Cadamuro, Davide and Goodsell, Mark and
                        Jaeckel, Joerg and Redondo, Javier and Ringwald, Andreas",
      title          = "{Wispy Cold Dark Matter}",
      journal        = "JCAP",
      volume         = "1206",
      year           = "2012",
      pages          = "013",
      doi            = "10.1088/1475-7516/2012/06/013",
      eprint         = "1201.5902",
      archivePrefix  = "arXiv",
      primaryClass   = "hep-ph",
      reportNumber   = "DESY-11-226, MPP-2011-140, CERN-PH-TH-2011-323,
                        IPPP-11-80, DCPT-11-160, --DCPT-11-160",
      SLACcitation   = "%%CITATION = ARXIV:1201.5902;%%"
}

@article{Graham:2015rva,
      author         = "Graham, Peter W. and Mardon, Jeremy and Rajendran,
                        Surjeet",
      title          = "{Vector Dark Matter from Inflationary Fluctuations}",
      journal        = "Phys. Rev.",
      volume         = "D93",
      year           = "2016",
      number         = "10",
      pages          = "103520",
      doi            = "10.1103/PhysRevD.93.103520",
      eprint         = "1504.02102",
      archivePrefix  = "arXiv",
      primaryClass   = "hep-ph",
      SLACcitation   = "%%CITATION = ARXIV:1504.02102;%%"
}

@article{Agrawal:2018vin,
      author         = "Agrawal, Prateek and Kitajima, Naoya and Reece, Matthew
                        and Sekiguchi, Toyokazu and Takahashi, Fuminobu",
      title          = "{Relic Abundance of Dark Photon Dark Matter}",
      journal        = "Phys. Lett.",
      volume         = "B801",
      year           = "2020",
      pages          = "135136",
      doi            = "10.1016/j.physletb.2019.135136",
      eprint         = "1810.07188",
      archivePrefix  = "arXiv",
      primaryClass   = "hep-ph",
      reportNumber   = "TU-1074,IPMU18-015,RESCEU-13/18,MIT-CTP/5066, TU-1074,
                        IPMU18-015, RESCEU-13/18, MIT-CTP/5066",
      SLACcitation   = "%%CITATION = ARXIV:1810.07188;%%"
}

@article{Co:2018lka,
      author         = "Co, Raymond T. and Pierce, Aaron and Zhang, Zhengkang and
                        Zhao, Yue",
      title          = "{Dark Photon Dark Matter Produced by Axion Oscillations}",
      journal        = "Phys. Rev.",
      volume         = "D99",
      year           = "2019",
      number         = "7",
      pages          = "075002",
      doi            = "10.1103/PhysRevD.99.075002",
      eprint         = "1810.07196",
      archivePrefix  = "arXiv",
      primaryClass   = "hep-ph",
      reportNumber   = "LCTP-18-21",
      SLACcitation   = "%%CITATION = ARXIV:1810.07196;%%"
}

@article{Nakai:2020cfw,
    author = "Nakai, Yuichiro and Namba, Ryo and Wang, Ziwei",
    title = "{Light Dark Photon Dark Matter from Inflation}",
    eprint = "2004.10743",
    archivePrefix = "arXiv",
    primaryClass = "hep-ph",
    doi = "10.1007/JHEP12(2020)170",
    journal = "JHEP",
    volume = "12",
    pages = "170",
    year = "2020"
}

@article{Dror:2018pdh,
    author = "Dror, Jeff A. and Harigaya, Keisuke and Narayan, Vijay",
    title = "{Parametric Resonance Production of Ultralight Vector Dark Matter}",
    eprint = "1810.07195",
    archivePrefix = "arXiv",
    primaryClass = "hep-ph",
    doi = "10.1103/PhysRevD.99.035036",
    journal = "Phys. Rev. D",
    volume = "99",
    number = "3",
    pages = "035036",
    year = "2019"
}

@article{Nakayama:2020rka,
    author = "Nakayama, Kazunori",
    title = "{Constraint on Vector Coherent Oscillation Dark Matter with Kinetic Function}",
    eprint = "2004.10036",
    archivePrefix = "arXiv",
    primaryClass = "hep-ph",
    reportNumber = "UT-20-04",
    doi = "10.1088/1475-7516/2020/08/033",
    journal = "JCAP",
    volume = "08",
    pages = "033",
    year = "2020"
}

@article{Ahmed:2020fhc,
    author = "Ahmed, Aqeel and Grzadkowski, Bohdan and Socha, Anna",
    title = "{Gravitational production of vector dark matter}",
    eprint = "2005.01766",
    archivePrefix = "arXiv",
    primaryClass = "hep-ph",
    doi = "10.1007/JHEP08(2020)059",
    journal = "JHEP",
    volume = "08",
    pages = "059",
    year = "2020"
}

@article{Sobol:2019xls,
      author         = "Sobol, O. O. and Gorbar, E. V. and Vilchinskii, S. I.",
      title          = "{Backreaction of electromagnetic fields and the Schwinger
                        effect in pseudoscalar inflation magnetogenesis}",
      journal        = "Phys. Rev.",
      volume         = "D100",
      year           = "2019",
      number         = "6",
      pages          = "063523",
      doi            = "10.1103/PhysRevD.100.063523",
      eprint         = "1907.10443",
      archivePrefix  = "arXiv",
      primaryClass   = "astro-ph.CO",
      SLACcitation   = "%%CITATION = ARXIV:1907.10443;%%"
}

@article{Sobol:2020frh,
    author = "Sobol, O. O. and Gorbar, E. V. and Momot, A. I. and Vilchinskii, S. I.",
    title = "{Schwinger production of scalar particles during and after inflation from the first principles}",
    eprint = "2004.12664",
    archivePrefix = "arXiv",
    primaryClass = "gr-qc",
    doi = "10.1103/PhysRevD.102.023506",
    journal = "Phys. Rev. D",
    volume = "102",
    number = "2",
    pages = "023506",
    year = "2020"
}

@article{Bertone:2018xtm,
      author         = "Bertone, Gianfranco and Tait, Tim, M. P.",
      title          = "{A new era in the search for dark matter}",
      journal        = "Nature",
      volume         = "562",
      year           = "2018",
      number         = "7725",
      pages          = "51-56",
      doi            = "10.1038/s41586-018-0542-z",
      eprint         = "1810.01668",
      archivePrefix  = "arXiv",
      primaryClass   = "astro-ph.CO",
      SLACcitation   = "%%CITATION = ARXIV:1810.01668;%%"
}

@article{Kolb:2020fwh,
    author = "Kolb, Edward W. and Long, Andrew J.",
    title = "{Completely Dark Photons from Gravitational Particle Production during the Inflationary Era}",
    eprint = "2009.03828",
    archivePrefix = "arXiv",
    primaryClass = "astro-ph.CO",
    doi = "10.1007/JHEP03(2021)283",
    journal = "JHEP",
    volume = "03",
    pages = "283",
    year = "2021"
}

@article{Salehian:2020asa,
    author = "Salehian, Borna and Gorji, Mohammad Ali and Firouzjahi, Hassan and Mukohyama, Shinji",
    title = "{Vector Dark Matter Production from Inflation with Symmetry Breaking}",
    eprint = "2010.04491",
    archivePrefix = "arXiv",
    primaryClass = "hep-ph",
    reportNumber = "YITP-20-123, IPMU20-0103",
    doi = "10.1103/PhysRevD.103.063526",
    journal = "Phys. Rev. D",
    volume = "103",
    number = "6",
    pages = "063526",
    year = "2021"
}

@article{Anber:2009ua,
    author = "Anber, Mohamed M. and Sorbo, Lorenzo",
    title = "{Naturally Inflating on Steep Potentials Through Electromagnetic Dissipation}",
    eprint = "0908.4089",
    archivePrefix = "arXiv",
    primaryClass = "hep-th",
    doi = "10.1103/PhysRevD.81.043534",
    journal = "Phys. Rev. D",
    volume = "81",
    pages = "043534",
    year = "2010"
}

@article{Anber:2006xt,
    author = "Anber, Mohamed M. and Sorbo, Lorenzo",
    title = "{N-flationary magnetic fields}",
    eprint = "astro-ph/0606534",
    archivePrefix = "arXiv",
    doi = "10.1088/1475-7516/2006/10/018",
    journal = "JCAP",
    volume = "10",
    pages = "018",
    year = "2006"
}

@article{Sobol:2021rlt,
    author = "Gorbar, E. V. and Schmitz, K. and Sobol, O. O. and Vilchinskii, S. I.",
    title = "{Gauge-field production during axion inflation in the gradient expansion formalism}",
    eprint = "2109.01651",
    archivePrefix = "arXiv",
    primaryClass = "hep-ph",
    reportNumber = "CERN-TH-2021-128",
    doi = "10.1103/PhysRevD.104.123504",
    journal = "Phys. Rev. D",
    volume = "104",
    number = "12",
    pages = "123504",
    year = "2021"
}

@article{Sobol:2023gwk,
    author = "von Eckardstein, Richard and Peloso, Marco and Schmitz, Kai and Sobol, Oleksandr and Sorbo, Lorenzo",
    title = "{Axion inflation in the strong-backreaction regime: decay of the Anber-Sorbo solution}",
    eprint = "2309.04254",
    archivePrefix = "arXiv",
    primaryClass = "hep-ph",
    reportNumber = "ACFI-T23-05, MS-TP-23-38",
    doi = "10.1007/JHEP11(2023)183",
    journal = "JHEP",
    volume = "11",
    pages = "183",
    year = "2023"
}

@article{Ema:2019yrd,
    author = "Ema, Yohei and Nakayama, Kazunori and Tang, Yong",
    title = "{Production of purely gravitational dark matter: the case of fermion and vector boson}",
    eprint = "1903.10973",
    archivePrefix = "arXiv",
    primaryClass = "hep-ph",
    reportNumber = "DESY-19-050, DESY 19-050, KEK-TH-2114, UT-19-04",
    doi = "10.1007/JHEP07(2019)060",
    journal = "JHEP",
    volume = "07",
    pages = "060",
    year = "2019"
}

@article{Firouzjahi:2020whk,
    author = "Firouzjahi, Hassan and Gorji, Mohammad Ali and Mukohyama, Shinji and Salehian, Borna",
    title = "{Dark photon dark matter from charged inflaton}",
    eprint = "2011.06324",
    archivePrefix = "arXiv",
    primaryClass = "hep-ph",
    doi = "10.1007/JHEP06(2021)050",
    journal = "JHEP",
    volume = "06",
    pages = "050",
    year = "2021"
}

@article{Cuissa:2018oiw,
    author = "Cuissa, Jose Roberto Canivete and Figueroa, Daniel G.",
    title = "{Lattice formulation of axion inflation. Application to preheating}",
    eprint = "1812.03132",
    archivePrefix = "arXiv",
    primaryClass = "astro-ph.CO",
    doi = "10.1088/1475-7516/2019/06/002",
    journal = "JCAP",
    volume = "06",
    pages = "002",
    year = "2019"
}

@article{Freese:1990rb,
    author = "Freese, Katherine and Frieman, Joshua A. and Olinto, Angela V.",
    title = "{Natural Inflation with Pseudo - Nambu-Goldstone Bosons}",
    reportNumber = "FERMILAB-PUB-90-177-A",
    doi = "10.1103/PhysRevLett.65.3233",
    journal = "Phys. Rev. Lett.",
    volume = "65",
    pages = "3233--3236",
    year = "1990"
}

@article{Adshead:2015pva,
    author = "Adshead, Peter and Giblin, John T. and Scully, Timothy R. and Sfakianakis, Evangelos I.",
    title = "{Gauge-preheating and the end of axion inflation}",
    eprint = "1502.06506",
    archivePrefix = "arXiv",
    primaryClass = "astro-ph.CO",
    doi = "10.1088/1475-7516/2015/12/034",
    journal = "JCAP",
    volume = "12",
    pages = "034",
    year = "2015"
}

@article{Starobinsky:1980te,
    author = "Starobinsky, Alexei A.",
    editor = "Khalatnikov, I. M. and Mineev, V. P.",
    title = "{A New Type of Isotropic Cosmological Models Without Singularity}",
    doi = "10.1016/0370-2693(80)90670-X",
    journal = "Phys. Lett. B",
    volume = "91",
    pages = "99--102",
    year = "1980"
}

@article{Mijic:1986iv,
    author = "Mijic, Milan B. and Morris, Michael S. and Suen, Wai-Mo",
    title = "{The R**2 Cosmology: Inflation Without a Phase Transition}",
    reportNumber = "CALT-68-1320",
    doi = "10.1103/PhysRevD.34.2934",
    journal = "Phys. Rev. D",
    volume = "34",
    pages = "2934",
    year = "1986"
}

@article{Maeda:1987xf,
    author = "Maeda, Kei-ichi",
    title = "{Inflation as a Transient Attractor in R**2 Cosmology}",
    reportNumber = "UTAP-60-87",
    doi = "10.1103/PhysRevD.37.858",
    journal = "Phys. Rev. D",
    volume = "37",
    pages = "858",
    year = "1988"
}

@article{Figueroa:2023oxc,
    author = "Figueroa, Daniel G. and Lizarraga, Joanes and Urio, Ander and Urrestilla, Jon",
    title = "{Strong Backreaction Regime in Axion Inflation}",
    eprint = "2303.17436",
    archivePrefix = "arXiv",
    primaryClass = "astro-ph.CO",
    doi = "10.1103/PhysRevLett.131.151003",
    journal = "Phys. Rev. Lett.",
    volume = "131",
    number = "15",
    pages = "151003",
    year = "2023"
}

@article{Barnaby:2010vf,
    author = "Barnaby, Neil and Peloso, Marco",
    title = "{Large Nongaussianity in Axion Inflation}",
    eprint = "1011.1500",
    archivePrefix = "arXiv",
    primaryClass = "hep-ph",
    reportNumber = "UMN-TH-2926-10",
    doi = "10.1103/PhysRevLett.106.181301",
    journal = "Phys. Rev. Lett.",
    volume = "106",
    pages = "181301",
    year = "2011"
}

@article{Figueroa:2024rkr,
    author = "Figueroa, Daniel G. and Lizarraga, Joanes and Loayza, Nicol{\'a}s and Urio, Ander and Urrestilla, Jon",
    title = "{Nonlinear dynamics of axion inflation: A detailed lattice study}",
    eprint = "2411.16368",
    archivePrefix = "arXiv",
    primaryClass = "astro-ph.CO",
    doi = "10.1103/PhysRevD.111.063545",
    journal = "Phys. Rev. D",
    volume = "111",
    number = "6",
    pages = "063545",
    year = "2025"
}

@article{ParticleDataGroup:2014cgo,
    author = "Olive, K. A. and others",
    collaboration = "Particle Data Group",
    title = "{Review of Particle Physics}",
    doi = "10.1088/1674-1137/38/9/090001",
    journal = "Chin. Phys. C",
    volume = "38",
    pages = "090001",
    year = "2014"
}

@article{DoF,
    author = "Husdal, Lars",
    title = "{On Effective Degrees of Freedom in the Early Universe}",
    eprint = "1609.04979",
    archivePrefix = "arXiv",
    primaryClass = "astro-ph.CO",
    doi = "10.3390/galaxies4040078",
    journal = "Galaxies",
    volume = "4",
    number = "4",
    pages = "78",
    year = "2016"
}

@article{Fixsen:2009ug,
    author = "Fixsen, D. J.",
    title = "{The Temperature of the Cosmic Microwave Background}",
    eprint = "0911.1955",
    archivePrefix = "arXiv",
    primaryClass = "astro-ph.CO",
    doi = "10.1088/0004-637X/707/2/916",
    journal = "Astrophys. J.",
    volume = "707",
    pages = "916--920",
    year = "2009"
}

@article{Baeza-Ballesteros:2026uao,
    author = "Baeza-Ballesteros, Jorge and Figueroa, Daniel G. and Florio, Adrien and Loayza, Nicol{\'a}s and Sattler, Franz R. and Torrent{\'\i}, Francisco and Urio, Ander",
    title = "{CosmoLattice 2.0}",
    eprint = "2607.24978",
    archivePrefix = "arXiv",
    primaryClass = "astro-ph.CO",
    reportNumber = "MS-TP-26-22, DESY-26-104",
    month = "7",
    year = "2026"
}

@article{LZ:2026axp,
    author = "Akerib, D. S. and others",
    collaboration = "LZ",
    title = "{Search for dark matter particle interactions in an extended nuclear recoil energy window with the LUX-ZEPLIN (LZ) experiment}",
    eprint = "2609.02823",
    archivePrefix = "arXiv",
    primaryClass = "hep-ex",
    doi = "10.17182/hepdata.182472.v1",
    month = "9",
    year = "2026"
}

@Book{CompAtomPhys,
  author    = "Bartschat, K.",
  title     = "Computational Atomic Physics",
  publisher = "Springer",
  year      = "1996"
}

@misc{COUL90,
author = "Barnett, A. R.",
title = {Ross {B}arnett's {C}oulomb and {B}essel functions},
howpublished = {Available at \url{https://www.fresco.org.uk/functions/barnett/index.htm} (2026/09/21)}
}

@article{Baudis:2025yva,
    author = "Baudis, Laura",
    title = "{Dark matter direct detection: status, results and future plans}",
    eprint = "2512.23039",
    archivePrefix = "arXiv",
    primaryClass = "astro-ph.IM",
    doi = "10.1088/1742-6596/3162/1/012007",
    journal = "J. Phys. Conf. Ser.",
    volume = "3162",
    number = "1",
    pages = "012007",
    year = "2025"
}

@inproceedings{Caputo:2026pdw,
    author = "Caputo, Andrea and Essig, Rouven",
    title = "{The Dark Photon: a 2026 Perspective}",
    booktitle = "Encyclopedia of Particle Physics",
    eprint = "2603.08430",
    archivePrefix = "arXiv",
    primaryClass = "hep-ph",
    month = "3",
    year = "2026"
}

@article{Barman:2025lvk,
    author = "Barman, Basabendu and Bernal, Nicol{\'a}s and Rubio, Javier",
    title = "{Two or three things particle physicists (mis)understand about (pre)heating}",
    eprint = "2503.19980",
    archivePrefix = "arXiv",
    primaryClass = "hep-ph",
    reportNumber = "IPARCOS-UCM-25-019",
    doi = "10.1016/j.nuclphysb.2025.116996",
    journal = "Nucl. Phys. B",
    volume = "1018",
    pages = "116996",
    year = "2025"
}

@article{Khlebnikov:1996mc,
    author = "Khlebnikov, S. Yu. and Tkachev, I. I.",
    title = "{Classical decay of inflaton}",
    eprint = "hep-ph/9603378",
    archivePrefix = "arXiv",
    reportNumber = "PURD-TH-96-02, OSU-TA-08-96",
    doi = "10.1103/PhysRevLett.77.219",
    journal = "Phys. Rev. Lett.",
    volume = "77",
    pages = "219--222",
    year = "1996"
}

@article{Dupac:2004pg,
    author = "Dupac, Xavier and Tauber, J.",
    title = "{Scanning strategy for mapping the Cosmic Microwave Background anisotropies with Planck}",
    eprint = "astro-ph/0409405",
    archivePrefix = "arXiv",
    doi = "10.1051/0004-6361:20041526",
    journal = "Astron. Astrophys.",
    volume = "430",
    pages = "363",
    year = "2005"
}

@article{Planck:2018vyg,
    author = "Aghanim, N. and others",
    collaboration = "Planck",
    title = "{Planck 2018 results. VI. Cosmological parameters}",
    eprint = "1807.06209",
    archivePrefix = "arXiv",
    primaryClass = "astro-ph.CO",
    doi = "10.1051/0004-6361/201833910",
    journal = "Astron. Astrophys.",
    volume = "641",
    pages = "A6",
    year = "2020",
    note = "[Erratum: Astron.Astrophys. 652, C4 (2021)]"
}

@article{Iarygina:2025ncl,
    author = "Iarygina, Oksana and Sfakianakis, Evangelos I. and Brandenburg, Axel",
    title = "{Schwinger effect in axion inflation on a lattice}",
    eprint = "2506.20538",
    archivePrefix = "arXiv",
    primaryClass = "astro-ph.CO",
    reportNumber = "NORDITA-2025-030",
    doi = "10.1103/1m23-ll9t",
    journal = "Phys. Rev. D",
    volume = "114",
    number = "6",
    pages = "063531",
    year = "2026"
}

@article{PencilCode:2020eyn,
    author = "Brandenburg, A. and others",
    collaboration = "Pencil Code",
    title = "{The Pencil Code, a modular MPI code for partial differential equations and particles: multipurpose and multiuser-maintained}",
    eprint = "2009.08231",
    archivePrefix = "arXiv",
    primaryClass = "astro-ph.IM",
    reportNumber = "NORDITA-2020-087",
    doi = "10.21105/joss.02807",
    journal = "J. Open Source Softw.",
    volume = "6",
    number = "58",
    pages = "2807",
    year = "2021"
}

@article{Lysenko:2025sdo,
    author = "Lysenko, A. V. and Sobol, O. O. and Vilchinskii, S. I.",
    title = "{Vector dark matter production during inflation in the gradient-expansion formalism}",
    eprint = "2509.24963",
    archivePrefix = "arXiv",
    primaryClass = "hep-ph",
    reportNumber = "MS-TP-25-28",
    doi = "10.1103/lpph-hrdr",
    journal = "Phys. Rev. D",
    volume = "113",
    number = "6",
    pages = "063525",
    year = "2026"
}

@article{Ferreira:2026klg,
    author = "Ferreira, Ricardo Z. and Notari, Alessio and Terente D{\'\i}az, Jos{\'e} Jaime",
    title = "{Axion Inflation with a Massive Abelian Gauge Field}",
    eprint = "2607.28593",
    archivePrefix = "arXiv",
    primaryClass = "astro-ph.CO",
    month = "7",
    year = "2026"
}

@article{Domcke:2026fjo,
    author = "Domcke, Valerie and Papageorgiou, Alexandros and Peloso, Marco and Sandner, Stefan",
    title = "{Axion Inflation: Perturbative control in the strong backreaction regime}",
    eprint = "2608.02498",
    archivePrefix = "arXiv",
    primaryClass = "astro-ph.CO",
    reportNumber = "CERN-TH-2026-181, LA-UR-26-26791",
    month = "8",
    year = "2026"
}

@article{Kofman:1997yn,
    author = "Kofman, Lev and Linde, Andrei D. and Starobinsky, Alexei A.",
    title = "{Towards the theory of reheating after inflation}",
    eprint = "hep-ph/9704452",
    archivePrefix = "arXiv",
    reportNumber = "IFA-97-28, SU-ITP-97-18",
    doi = "10.1103/PhysRevD.56.3258",
    journal = "Phys. Rev. D",
    volume = "56",
    pages = "3258--3295",
    year = "1997"
}

@article{Nagata:2026pbj,
    author = "Nagata, Natsumi and Yanagida, Tsutomu T.",
    title = "{Asymmetric Inelastic Dark Matter and the LUX-ZEPLIN event}",
    eprint = "2609.18564",
    archivePrefix = "arXiv",
    primaryClass = "hep-ph",
    month = "9",
    year = "2026"
}

\end{document}